\documentstyle[12pt,epsf]{article}
\newlength{\pubnumber} 

\catcode`\@=11
\@addtoreset{equation}{section}

\def\section{\@startsection{section}{1}{\z@}{3.5ex plus 1ex minus .2ex}
 {2.3ex plus .2ex}{\large\bf}}
\def\subsection{\@startsection{subsection}{2}{\z@}{2.3ex plus .2ex}
 {2.3ex plus .2ex}{\bf}}

\newcommand{\cc}[2]{c{#1\atopwithdelims[]#2}}
\input epsf

\begin{document}

\begin{titlepage}
\samepage{
\setcounter{page}{1}
\rightline{LTH--646}
\rightline{OUTP--05--03P}
\rightline{\tt hep-th/0504016}
\rightline{March 2005}
\vfill
\begin{center}
 {\Large \bf  Moduli Fixing in Realistic String Vacua}
\vfill
\vfill {\large Alon E. Faraggi\footnote{faraggi@thphys.ox.ac.uk}}\\
\vspace{.12in}
{\it         Theoretical Physics Department, University of Oxford,
            Oxford, OX1 3NP, UK\\}
\vspace{.025in}

{\it   and}

\vspace{.025in}
{\it         Department of Mathematical Sciences, University of Liverpool,
            Liverpool, L69 3BX, UK\\}

\end{center}
\vfill
\begin{abstract}
I demonstrate the existence of quasi--realistic heterotic--string
models in which all the untwisted K\"ahler and complex structure 
moduli, as well as all of the twisted sectors moduli,
are projected out by the generalized GSO projections.
I discuss the conditions and characteristics of the models
that produce this result.
The existence of such models offers a novel perspective
on the realization of extra dimensions in string theory.
In this view, while the geometrical picture provides a useful
mean to classify string vacua, in the phenomenologically viable cases
there is no physical realization of extra dimensions.
The models under consideration correspond to
$Z_2\times Z_2$ orbifolds of six dimensional tori,
plus additional identifications by internal shifts
and twists. The special property of the $Z_2\times Z_2$
orbifold is that it may act on the compactified dimensions
as real, rather than complex, dimensions. This property enables
an asymmetric projection on all six internal coordinates,
which enables the projection of all the untwisted K\"ahler and
complex structure moduli. 
\end{abstract}
\smallskip}
\end{titlepage}

\setcounter{footnote}{0}

\def\beq{\begin{equation}}
\def\eeq{\end{equation}}
\def\beqn{\begin{eqnarray}}
\def\eeqn{\end{eqnarray}}

\def\no{\noindent }
\def\nolabel{\nonumber }
\def\ie{{\it i.e.}}
\def\eg{{\it e.g.}}
\def\half{{\textstyle{1\over 2}}}
\def\third{{\textstyle {1\over3}}}
\def\quarter{{\textstyle {1\over4}}}
\def\sixth{{\textstyle {1\over6}}}
\def\m{{\tt -}}
\def\p{{\tt +}}

\def\Tr{{\rm Tr}\, }
\def\tr{{\rm tr}\, }

\def\slash#1{#1\hskip-6pt/\hskip6pt}
\def\slk{\slash{k}}
\def\GeV{\,{\rm GeV}}
\def\TeV{\,{\rm TeV}}
\def\y{\,{\rm y}}
\def\SM{Standard--Model }
\def\SUSY{supersymmetry }
\def\SSSM{supersymmetric standard model}
\def\vev#1{\left\langle #1\right\rangle}
\def\l{\langle}
\def\r{\rangle}
\def\o#1{\frac{1}{#1}}

\def\Htw{{\tilde H}}
\def\chibar{{\overline{\chi}}}
\def\qbar{{\overline{q}}}
\def\ibar{{\overline{\imath}}}
\def\jbar{{\overline{\jmath}}}
\def\Hbar{{\overline{H}}}
\def\Qbar{{\overline{Q}}}
\def\abar{{\overline{a}}}
\def\alphabar{{\overline{\alpha}}}
\def\betabar{{\overline{\beta}}}
\def\tautwo{{ \tau_2 }}
\def\thetatwo{{ \vartheta_2 }}
\def\thetathree{{ \vartheta_3 }}
\def\thetafour{{ \vartheta_4 }}
\def\ttwo{{\vartheta_2}}
\def\tthree{{\vartheta_3}}
\def\tfour{{\vartheta_4}}
\def\ti{{\vartheta_i}}
\def\tj{{\vartheta_j}}
\def\tk{{\vartheta_k}}
\def\calF{{\cal F}}
\def\smallmatrix#1#2#3#4{{ {{#1}~{#2}\choose{#3}~{#4}} }}
\def\ab{{\alpha\beta}}
\def\Minv{{ (M^{-1}_\ab)_{ij} }}
\def\bone{{\bf 1}}
\def\ii{{(i)}}
\def\V{{\bf V}}
\def\N{{\bf N}}
\def\lv{\vert 0\rangle_L} 
\def\rv{\vert 0\rangle_R} 

\def\b{{\bf b}}
\def\S{{\bf S}}
\def\X{{\bf X}}
\def\I{{\bf I}}
\def\mb{{\mathbf b}}
\def\mS{{\mathbf S}}
\def\mX{{\mathbf X}}
\def\mI{{\mathbf I}}
\def\balpha{{\mathbf \alpha}}
\def\bbeta{{\mathbf \beta}}
\def\bgamma{{\mathbf \gamma}}
\def\bxi{{\mathbf \xi}}

\def\t#1#2{{ \Theta\left\lbrack \matrix{ {#1}\cr {#2}\cr }\right\rbrack }}
\def\C#1#2{{ C\left\lbrack \matrix{ {#1}\cr {#2}\cr }\right\rbrack }}
\def\tp#1#2{{ \Theta'\left\lbrack \matrix{ {#1}\cr {#2}\cr }\right\rbrack }}
\def\tpp#1#2{{ \Theta''\left\lbrack \matrix{ {#1}\cr {#2}\cr }\right\rbrack }}
\def\l{\langle}
\def\r{\rangle}


\def\inbar{\,\vrule height1.5ex width.4pt depth0pt}

\def\IC{\relax\hbox{$\inbar\kern-.3em{\rm C}$}}
\def\IQ{\relax\hbox{$\inbar\kern-.3em{\rm Q}$}}
\def\IR{\relax{\rm I\kern-.18em R}}
 \font\cmss=cmss10 \font\cmsss=cmss10 at 7pt
\def\IZ{\relax\ifmmode\mathchoice
 {\hbox{\cmss Z\kern-.4em Z}}{\hbox{\cmss Z\kern-.4em Z}}
 {\lower.9pt\hbox{\cmsss Z\kern-.4em Z}}
 {\lower1.2pt\hbox{\cmsss Z\kern-.4em Z}}\else{\cmss Z\kern-.4em Z}\fi}

\def\AEF{A.E. Faraggi}
\def\CMP#1#2#3{{\it Comm.\ Math.\ Phys.}\/ {\bf #1} (#2) #3}
\def\JHEP#1#2#3{JHEP {\textbf #1}, (#2) #3}
\def\NPB#1#2#3{{\it Nucl.\ Phys.}\/ {\bf B#1} (#2) #3}
\def\PLB#1#2#3{{\it Phys.\ Lett.}\/ {\bf B#1} (#2) #3}
\def\PRD#1#2#3{{\it Phys.\ Rev.}\/ {\bf D#1} (#2) #3}
\def\PRL#1#2#3{{\it Phys.\ Rev.\ Lett.}\/ {\bf #1} (#2) #3}
\def\PRT#1#2#3{{\it Phys.\ Rep.}\/ {\bf#1} (#2) #3}
\def\MODA#1#2#3{{\it Mod.\ Phys.\ Lett.}\/ {\bf A#1} (#2) #3}
\def\IJMP#1#2#3{{\it Int.\ J.\ Mod.\ Phys.}\/ {\bf A#1} (#2) #3}
\def\nuvc#1#2#3{{\it Nuovo Cimento}\/ {\bf #1A} (#2) #3}
\def\RPP#1#2#3{{\it Rept.\ Prog.\ Phys.}\/ {\bf #1} (#2) #3}
\def\etal{{\it et al\/}}

\hyphenation{su-per-sym-met-ric non-su-per-sym-met-ric}
\hyphenation{space-time-super-sym-met-ric}
\hyphenation{mod-u-lar mod-u-lar--in-var-i-ant}


\setcounter{footnote}{0}
\section{Introduction}
String theory continues to serve as the most compelling framework
for the unification of all the fundamental matter and interactions.
Amazingly, preservation of some classical string symmetries necessitates
the introduction of a fixed number of degrees of freedom,
beyond the observable ones. One interpretation of these
additional degrees of freedom is as extra space--time
dimensions, which clearly is in contradiction  with the
observed facts and diminishes from the appeal of string
theory. Thus, the need arises to hide the additional dimensions
by compactifying the superstring on a six dimensional compact
manifold. This in turn leads to the problem of fixing the
parameters of the extra dimensions in such a way that it does
not conflict with contemporary experimental limits.
Additionally, the extra dimensions give rise to light
scalar fields whose VEV parameterizes the shape and size
of the extra dimensions. 
This is typically dubbed as the moduli stabilization problem, and
preoccupies much of the activity in string theory \cite{recentpaper}
ever since the initial realization of its potential relevance for
particle phenomenology \cite{heterotic,canwit}.

A particular class of string compactifications that exhibit
promising phenomenological properties are the so--called
free fermionic heterotic--string models 
\cite{costas,revamp,fny,alr,nahe,eu,top,cslm,others,pairings}. 
These models are 
related to $Z_2\times Z_2$ orbifold compactifications \cite{foc},
and many of their appealing phenomenological properties
are rooted in the underlying $Z_2\times Z_2$ orbifold structure \cite{foc}.
Recently, a classification of this class of models was initiated
and it was revealed that in a large class of models the
reduction to three generations necessitates the utilization
of a fully asymmetric shift \cite{fknr}. This indicates that the geometrical
objects underlying these models do not correspond to
the standard complex geometries that have been the most
prevalent in the literature. 

Furthermore, the utilization of the asymmetric shift has important
bearing on the issue of moduli fixing in string theory
\cite{asymmetric}, and in this class
of string compactifications and in particular.
The operation of an asymmetric shift can only occur at 
special points in the moduli space, and results in fixation
of the moduli at the special points. It is therefore of interest
to investigate this aspect in greater detail.
In the context of the free fermionic models the
marginal operators that correspond to the moduli
deformations correspond to the inclusion of 
Thirring interactions among the world--sheet fermions
\cite{fff,bagger,thirring,lny}.
The allowed world--sheet Thirring interactions must 
be invariant under the GSO projections that are induced 
by the boundary condition basis vectors that define the
free fermionic string models. Thus, depending on the assignment
of boundary conditions in specific models, some of
the world--sheet Thirring interactions are forbidden, 
and their corresponding moduli are projected from the
physical spectrum. In this manner the GSO projection
induced by the boundary condition basis vectors acts
as a moduli fixing mechanism. 

In this paper I discuss the existence of
quasi--realistic three generation free fermionic
models, in which all the untwisted K\"ahler and complex structure
moduli are projected out from the low energy effective
field theory. This is a striking result with crucial
implications for the premise of extra dimensions in
physical string vacua. This result indicates that the
interpretation of the additional degrees of freedom needed
to obtain a conformally invariant string theory as additional
continuous spatial dimensions beyond the observable ones, may in fact not
be realized in the phenomenologically relevant cases. Furthermore,
the fact that the projection is achieved by an asymmetric orbifold
action indicates that the treatment of the moduli problem in the
low energy effective supergravity theory is inadequate.
The reason being that the effective supergravity theories
relate to the additional dimensions as classical geometries,
and hence necessarily as left--right symmetric. On the other
hand the basic property of string theory is that it allows 
for the left--right world--sheet asymmetry, which in the
context of the phenomenological free fermionic models is
instrumental in fixing the moduli.

\section{General structure of free fermionic models}\label{rffm}

In the free fermionic formulation \cite{fff} of the heterotic string
in four dimensions all the world--sheet
degrees of freedom  required to cancel
the conformal anomaly are represented in terms of free fermions
propagating on the string world--sheet.
In the light--cone gauge the world--sheet field content consists
of two transverse left-- and right--moving space--time coordinate bosons,
$X_{1,2}^\mu$ and ${\bar X}_{1,2}^\mu$,
and their left--moving fermionic superpartners $\psi^\mu_{1,2}$,
and additional 62 purely internal
Majorana--Weyl fermions, of which 18 are left--moving,
$\chi^{I}$, and 44 are right--moving, $\phi^a$.
In the supersymmetric sector the world--sheet supersymmetry is realized
non--linearly and the world--sheet supercurrent is given by
\begin{equation}
T_F=\psi^\mu\partial X_\mu+f_{IJK}\chi^I\chi^J\chi^K,
\label{supercurrent}
\end{equation}
where $f_{IJK}$ are the structure constants of a semi--simple
Lie group of dimension 18. The $\chi^{I}~(I=1,\cdots,18)$
world--sheet fermions transform in the adjoint representation of
the Lie group. In the realistic free fermionic models the Lie group is
$SU(2)^6$. The $\chi^I~{I=1,\cdots,18}$ transform in the adjoint
representation of $SU(2)^6$, and are denoted by
$\chi^I,~y^I,~\omega^I~(I=1,\cdots,6)$.
Under parallel transport around a noncontractible loop on the toroidal
world--sheet the fermionic fields pick up a phase
\begin{equation}
f~\rightarrow~-{\rm e}^{i\pi\alpha(f)}f~.
\label{fermionphase}
\end{equation}
The minus sign is conventional and $\alpha(f)\in(-1,+1]$.
A model in this construction \cite{fff}
is defined by a set of boundary conditions basis vectors
and by a choice of generalized GSO projection coefficients, which
satisfy the one--loop modular invariance constraints. The boundary
conditions basis vectors ${\bf b}_k$ span a finite additive group
\begin{equation}
\Xi={\sum_k}n_i {b}_i
\label{additivegroup}
\end{equation}
where $n_i=0,\cdots,{{N_{z_i}}-1}$.
The physical massless states in the Hilbert space of a given sector
$\alpha\in{\Xi}$ are then obtained by acting on the vacuum state of
that sector with the world-sheet bosonic and fermionic mode operators,
with frequencies $\nu_f$, $\nu_{f^*}$ and
by subsequently applying the generalized GSO projections,
\begin{equation}
\left\{e^{i\pi({b_i}F_\alpha)}-
{\delta_\alpha}c^*\left(\matrix{\alpha\cr
                 b_i\cr}\right)\right\}\vert{s}\rangle=0
\label{gsoprojections}
\end{equation}
with
\begin{equation}
(b_i{F_\alpha})\equiv\{\sum_{real+complex\atop{left}}-
\sum_{real+complex\atop{right}}\}(b_i(f)F_\alpha(f)),
\label{lorentzproduct}
\end{equation}
where $F_\alpha(f)$ is a fermion number operator counting each mode of
$f$ once (and if $f$ is complex, $f^*$ minus once). For periodic
complex fermions [{\it i.e.} for $\alpha(f)=1)$]
the vacuum is a spinor in order to represent
the Clifford algebra of the corresponding zero modes.
For each periodic complex fermion $f$,
there are two degenerate vacua $\vert{+}\rangle$, $\vert{-}\rangle$,
annihilated by the zero modes $f_0$ and $f^*_0$ and with fermion
number $F(f)=0,-1$ respectively. In Eq. (\ref{gsoprojections}),
$\delta_\alpha=-1$ if $\psi^\mu$ is periodic in the sector $\alpha$,
and $\delta_\alpha=+1$ if $\psi^\mu$ is antiperiodic in the sector $\alpha$.
Each complex world--sheet fermion $f$ gives rise to a $U(1)$ $ff^*$
current in the Cartan sub--algebra of the four dimensional gauge group, 
with the charge given by:
\beq
Q(f)~=~{1\over 2}\alpha(f) + F(f)~.
\label{qoff}
\eeq
Alternatively, a real left--moving fermion may be combined with 
a real right--moving fermion to form an Ising model operator
\cite{kln,fm}, in which case the $U(1)$ generators are projected
out, and the rank of the four dimensional gauge group is reduced.
This distinction between complex and real world--sheet fermion
has important bearing on the assignment of 
asymmetric versus symmetric boundary conditions, and hence
on the issue of moduli fixing in the $Z_2\times Z_2$ fermionic
models. 

The ${\delta_\alpha}$ in eq. (\ref{gsoprojections}) is a space--time
spin statistics factor, equal to $+1$ for space--time bosons and $-1$
for space--time fermions, and is determined by the boundary conditions
of the space--time fermions $\psi^\mu_{1,2}$ in the sector $\alpha$,
{\it i.e.}, 
$$\delta_\alpha= {\rm e}^{i\pi \alpha(\psi^\mu)}.$$
The $$c^*\left(\matrix{\alpha\cr b_i\cr}\right)$$ are free phases in
the one--loop string partition function. For the Neveu--Schwarz untwisted  
sector we have the general result that
$$\delta_{NS} c^*\left(\matrix{NS\cr b\cr}\right)=\delta_{b}.$$
Hence, the free phases do not play a significant role in the
determination of the untwisted moduli. The existence of the
untwisted moduli in a model depends solely on the boundary conditions.
The free phases of the models will therefore be suppressed in the following.

The boundary condition basis defining a typical realistic free
fermionic heterotic string model is constructed in two stages. The
first stage consists of the NAHE set, which is a set of five
boundary condition basis vectors, $\{ 1 ,S,b_1,b_2,b_3\}$
\cite{costas,nahe,foc,pairings}. The gauge group induced by the NAHE set is
${\rm SO} (10)\times {\rm SO}(6)^3\times {\rm E}_8$ with ${ N}=1$
supersymmetry. The space-time vector bosons that generate the
gauge group arise from the Neveu--Schwarz sector and from the
sector $\xi_2\equiv 1+b_1+b_2+b_3$. The Neveu-Schwarz sector
produces the generators of ${\rm SO}(10)\times {\rm SO}(6)^3\times
{\rm SO}(16)$. The $\xi_2$-sector produces the spinorial 128 of
SO(16) and completes the hidden gauge group to ${\rm E}_8$. The
NAHE set divides the internal world-sheet fermions in the
following way:
${\bar\phi}^{1,\cdots,8}$ generate the hidden ${\rm E}_8$ gauge group,
${\bar\psi}^{1,\cdots,5}$ generate the SO(10) gauge group, and
$\{{\bar y}^{3,\cdots,6},{\bar\eta}^1\}$,
$\{{\bar y}^1,{\bar y}^2,{\bar\omega}^5,{\bar\omega}^6,{\bar\eta}^2\}$,
$\{{\bar\omega}^{1,\cdots,4},{\bar\eta}^3\}$
generate the three horizontal ${\rm SO}(6)$ symmetries.
The left-moving $\{y,\omega\}$ states are divided into
$\{{y}^{3,\cdots,6}\}$,
$\{{y}^1,{y}^2,{\omega}^5,{\omega}^6\}$,
$\{{\omega}^{1,\cdots,4}\}$ and
$\chi^{12}$, $\chi^{34}$, $\chi^{56}$ generate the left--moving
${N}=2$ world--sheet supersymmetry.
At the level of the NAHE set the sectors $b_1$, $b_2$ and $b_3$
produce 48 multiplets, 16 from each, in the $16$
representation of SO(10). The states from the sectors $b_j$ are
singlets of the hidden ${\rm E}_8$ gauge group and transform under
the horizontal ${\rm SO}(6)_j$ $(j=1,2,3)$ symmetries. This
structure is common to a large set of realistic free fermionic models.

The second stage of the construction consists of adding to the
NAHE set three (or four) additional basis vectors.
These additional vectors reduce the number of generations
to three, one from each of the sectors $b_1$, $b_2$ and $b_3$,
and simultaneously break the four dimensional gauge group.
The assignment of boundary conditions to
$\{{\bar\psi}^{1,\cdots,5}\}$ breaks SO(10) to one of its subgroups
\cite{cslm}.
Similarly, the hidden ${\rm E}_8$ symmetry is broken to one of its
subgroups, and the flavor ${\rm SO}(6)^3$ symmetries are broken
to $U(1)^n$, with $3\le n\le9$.
For details and phenomenological studies of
these three generation string models interested readers are 
referred to the original literature and review articles \cite{reviewsp}.

The correspondence of the free fermionic models
with the orbifold construction \cite{dhvw} is facilitated
by extending the NAHE set, $\{ 1,S,b_1,b_2,b_3\}$, by at least
one additional boundary condition basis vector \cite{foc}
\beq
\xi_1=(0,\cdots,0\vert{\underbrace{1,\cdots,1}_{{\bar\psi^{1,\cdots,5}},
{\bar\eta^{1,2,3}}}},0,\cdots,0)~.
\label{vectorx1}
\eeq
With a suitable choice of the GSO projection coefficients the
model possesses an ${\rm SO}(4)^3\times {\rm E}_6\times {\rm U}(1)^2
\times {\rm E}_8$ gauge group
and ${ N}=1$ space-time supersymmetry. The matter fields
include 24 generations in the 27 representation of
${\rm E}_6$, eight from each of the sectors $b_1\oplus b_1+\xi_1$,
$b_2\oplus b_2+\xi_1$ and $b_3\oplus b_3+\xi_1$.
Three additional 27 and $\overline{27}$ pairs are obtained
from the Neveu-Schwarz (NS) $\oplus~\xi_1$ sector, that correspond
to the untwisted sector of the orbifold models.

To construct the model in the orbifold formulation one starts
with the compactification on a torus with nontrivial background
fields \cite{narain}.
The subset of basis vectors
\beq
\{ 1,S,\xi_1,\xi_2\}
\label{neq4set}
\eeq
generates a toroidally-compactified model with ${N}=4$ space-time
supersymmetry and ${\rm SO}(12)\times {\rm E}_8\times {\rm E}_8$ gauge group.
The same model is obtained in the geometric (bosonic) language
by tuning the background fields to the values corresponding to
the SO(12) lattice. The
metric of the six-dimensional compactified
manifold is then the Cartan matrix of SO(12),
while the antisymmetric tensor is given by
\beq
B_{ij}=\cases{
G_{ij}&;\ $i>j$,\cr
0&;\ $i=j$,\cr
-G_{ij}&;\ $i<j$.\cr}
\label{bso12}
\eeq
When all the radii of the six-dimensional compactified
manifold are fixed at $R_I=\sqrt2$, it is seen that the
left- and right-moving momenta
$
P^I_{R,L}=[m_i-{\frac{1}{2}}(B_{ij}{\pm}G_{ij})n_j]{e_i^I}^*
$
reproduce the massless root vectors in the lattice of
SO(12). Here $e^i=\{e_i^I\}$ are six linearly-independent
vielbeins normalized so that $(e_i)^2=2$.
The ${e_i^I}^*$ are dual to the $e_i$, with
$e_i^*\cdot e_j=\delta_{ij}$.

Adding the two basis vectors $b_1$ and $b_2$ to the set
(\ref{neq4set}) corresponds to the ${Z}_2\times {Z}_2$
orbifold model with standard embedding.
Starting from the $N=4$ model with ${\rm SO}(12)\times
{\rm E}_8\times {\rm E}_8$
symmetry, and applying the ${Z}_2\times {Z}_2$
twist on the
internal coordinates, reproduces
the spectrum of the free-fermion model
with the six-dimensional basis set
$\{ 1,S,\xi_1,\xi_2,b_1,b_2\}$.
The Euler characteristic of this model is 48 with $h_{11}=27$ and
$h_{21}=3$.

The effect of the additional basis vector $\xi_1$ of eq.
(\ref{vectorx1}), is to separate the gauge degrees of freedom, spanned by
the world-sheet fermions $\{{\bar\psi}^{1,\cdots,5},
{\bar\eta}^{1},{\bar\eta}^{2},{\bar\eta}^{3},{\bar\phi}^{1,\cdots,8}\}$,
from the internal compactified degrees of freedom
$\{y,\omega\vert {\bar y},{\bar\omega}\}^{1,\cdots,6}$.
In the realistic free fermionic models \cite{cslm,foc}
this is achieved by the vector $2\gamma$ \cite{foc}, with
\beq
2\gamma=(0,\cdots,0\vert{\underbrace{1,\cdots,1}_{{\bar\psi^{1,\cdots,5}},
{\bar\eta^{1,2,3}} {\bar\phi}^{1,\cdots,4}} },0,\cdots,0)~,
\label{vector2gamma}
\eeq
which breaks the ${\rm E}_8\times {\rm E}_8$ symmetry to ${\rm SO}(16)\times
{\rm SO}(16)$.
The ${Z}_2\times {Z}_2$ twist induced by $b_1$ and $b_2$
breaks the gauge symmetry to
${\rm SO}(4)^3\times {\rm SO}(10)\times {\rm U}(1)^3\times {\rm SO}(16)$.
The orbifold still yields a model with 24 generations,
eight from each twisted sector,
but now the generations are in the chiral 16 representation
of SO(10), rather than in the 27 of ${\rm E}_6$. The same model can
be realized with the set
$\{ 1,S,\xi_1,\xi_2,b_1,b_2\}$,
by projecting out the $16\oplus{\overline{16}}$
from the $\xi_1$-sector taking
\beq\label{changec}
\cc{\xi_1}{\xi_2} \rightarrow -\cc{\xi_1}{\xi_2}.
\label{gsophasexi1xi2}
\eeq
This choice also projects out the massless vector bosons in the
128 of SO(16) in the hidden-sector ${\rm E}_8$ gauge group, thereby
breaking the ${\rm E}_6\times {\rm E}_8$ symmetry to
${\rm SO}(10)\times {\rm U}(1)\times {\rm SO}(16)$.
However, the ${Z}_2\times {Z}_2$ twist acts identically
in the two models, and their physical characteristics
differ only due to the discrete torsion eq. (\ref{changec}).
This analysis confirms that the $Z_2\times Z_2$
orbifold on the $SO(12)$ lattice is at the core of the realistic
free fermionic models.

The set of real fermions $\{y,\omega\vert {\bar y},{\bar\omega}\}$
correspond to the six dimensional compactified coordinates in a 
bosonic formulation. The assignment of boundary conditions
to this set of internal fermions therefore correspond
to the action on the internal six dimensional compactified
manifold. Consequently, the assignment
of boundary conditions to this set of real fermions
determines many of the phenomenological properties
of the low energy spectrum and effective field theory. 
Examples include the determination of the number
of light generations \cite{nahe,pairings}; the stringy doublet--triplet
splitting mechanism \cite{ps}; and the top--bottom--quark Yukawa
coupling selection mechanism \cite{topbottomyuk}.
In the last two cases, it is the left--right asymmetric boundary
conditions that enables the doublet--triplet splitting, as well as
the Yukawa coupling selection mechanism \cite{ps,topbottomyuk}.
In this paper I discuss how the asymmetric assignment of boundary
conditions to the set of internal world--sheet
fermions $\{y,\omega\vert {\bar y},{\bar\omega}\}$, also
fixes the untwisted moduli. 
It is therefore found that the same condition, 
namely the left--right orbifold asymmetry,
is the one that plays the crucial role, both in the determination
of the physical properties mentioned above, as well as in
the fixation of the moduli parameters. 

The symmetric versus asymmetric orbifold action is determined
in the free fermionic models by the pairing of the left--right
real internal fermions from the set
$\{y,\omega\vert {\bar y},{\bar\omega}\}$,
into real pairs, that pair a left--moving fermion
with a right--moving fermion, versus complex pairs,
that combine left--left, or right--right, moving fermions.
The real pairs preserve the left--right symmetry,
whereas the complex pairs allow for the assignment
of asymmetric boundary conditions, that correspond
to asymmetric action on the compactified bosonic
coordinates. 

The reduction of the number of generations to three is illustrated
in tables [\ref{m278}] and [\ref{symmetric3gen}].

\begin{eqnarray}
 &\begin{tabular}{c|c|ccc|c|ccc|c}
 ~ & $\psi^\mu$ & $\chi^{12}$ & $\chi^{34}$ & $\chi^{56}$ &
        $\bar{\psi}^{1,...,5} $ &
        $\bar{\eta}^1 $&
        $\bar{\eta}^2 $&
        $\bar{\eta}^3 $&
        $\bar{\phi}^{1,...,8} $\\
\hline
\hline
  ${\alpha}$  &  1 & 1&0&0 & 1~1~1~0~0 & 1 & 0 & 0 & 1~1~0~0~0~0~0~0 \\
  ${\beta}$   &  1 & 0&1&0 & 1~1~1~0~0 & 0 & 1 & 0 & 1~1~0~0~0~0~0~0 \\
  ${\gamma}$  &  1 & 0&0&1 &
		${1\over2}$~${1\over2}$~${1\over2}$~${1\over2}$~${1\over2}$
	      & ${1\over2}$ & ${1\over2}$ & ${1\over2}$ &
                ${1\over2}$~${1\over2}$~${1\over2}$~${1\over2}$~1~0~0~0 \\
\end{tabular}
   \nonumber\\
   ~  &  ~ \nonumber\\
   ~  &  ~ \nonumber\\
     &\begin{tabular}{c|c|c|c}
 ~&   $y^3{\bar y}^3$
      $y^4{\bar y}^4$
      $y^5{\bar y}^5$
      ${y}^6{\bar y}^6$
  &   $y^1{\bar y}^1$
      $y^2{\bar y}^2$
      $\omega^5{\bar\omega}^5$
      ${\omega}^6{\bar\omega}^6$
  &   $\omega^1{\bar\omega}^1$
      $\omega^2{\bar\omega}^2$
      $\omega^3{\bar\omega}^3$
      ${\omega}^4{\bar\omega}^4$ \\
\hline
\hline
$\alpha$ & 1 ~~~ 0 ~~~ 0 ~~~ 1  & 0 ~~~ 0 ~~~ 1 ~~~ 0  & 0 ~~~ 0 ~~~ 0 ~~~ 1
\\
$\beta$  & 0 ~~~ 0 ~~~ 0 ~~~ 1  & 0 ~~~ 1 ~~~ 1 ~~~ 0  & 1 ~~~ 0 ~~~ 0 ~~~ 0
\\
$\gamma$ & 1 ~~~ 1 ~~~ 0 ~~~ 0  & 1 ~~~ 0 ~~~ 0 ~~~ 0  & 0 ~~~ 1 ~~~ 0 ~~~ 0
\\
\end{tabular}
\label{symmetric3gen}
\end{eqnarray}

\begin{eqnarray}
 &\begin{tabular}{c|c|ccc|c|ccc|c}
 ~ & $\psi^\mu$ & $\chi^{12}$ & $\chi^{34}$ & $\chi^{56}$ &
        $\bar{\psi}^{1,...,5} $ &
        $\bar{\eta}^1 $&
        $\bar{\eta}^2 $&
        $\bar{\eta}^3 $&
        $\bar{\phi}^{1,...,8} $\\
\hline
\hline
  ${\alpha}$  &  0 & 0&0&0 & 1~1~1~0~0 & 0 & 0 & 0 & 1~1~1~1~0~0~0~0 \\
  ${\beta}$   &  0 & 0&0&0 & 1~1~1~0~0 & 0 & 0 & 0 & 1~1~1~1~0~0~0~0 \\
  ${\gamma}$  &  0 & 0&0&0 &
		${1\over2}$~${1\over2}$~${1\over2}$~${1\over2}$~${1\over2}$
	      & ${1\over2}$ & ${1\over2}$ & ${1\over2}$ &
                ${1\over2}$~0~1~1~${1\over2}$~${1\over2}$~${1\over2}$~0 \\
\end{tabular}
   \nonumber\\
   ~  &  ~ \nonumber\\
   ~  &  ~ \nonumber\\
     &\begin{tabular}{c|c|c|c}
 ~&   $y^3{y}^6$
      $y^4{\bar y}^4$
      $y^5{\bar y}^5$
      ${\bar y}^3{\bar y}^6$
  &   $y^1{\omega}^5$
      $y^2{\bar y}^2$
      $\omega^6{\bar\omega}^6$
      ${\bar y}^1{\bar\omega}^5$
  &   $\omega^2{\omega}^4$
      $\omega^1{\bar\omega}^1$
      $\omega^3{\bar\omega}^3$
      ${\bar\omega}^2{\bar\omega}^4$ \\
\hline
\hline
$\alpha$ & 1 ~~~ 0 ~~~ 0 ~~~ 0  & 0 ~~~ 0 ~~~ 1 ~~~ 1  & 0 ~~~ 0 ~~~ 1 ~~~ 1
\\
$\beta$  & 0 ~~~ 0 ~~~ 1 ~~~ 1  & 1 ~~~ 0 ~~~ 0 ~~~ 0  & 0 ~~~ 1 ~~~ 0 ~~~ 1
\\
$\gamma$ & 0 ~~~ 1 ~~~ 0 ~~~ 1  & 0 ~~~ 1 ~~~ 0 ~~~ 1  & 1 ~~~ 0 ~~~ 0 ~~~ 0
\\
\end{tabular}
\label{m278}
\end{eqnarray}
together with a consistent choice of one--loop GSO phases \cite{eu}.
Both models produce three generations and the $SO(10)$ GUT gauge
group is broken to $SU(3)\times SU(2)\times U(1)^2$. The 
model of table [\ref{symmetric3gen}] produces three pairs of 
$SU(3)$ color triplets from the untwisted sector, whereas that of
table [\ref{m278}] produces the corresponding three pairs of electroweak
Higgs doublets \cite{eu,ps}. The set of boundary conditions in table
[\ref{symmetric3gen}] is symmetric between the left and right movers,
whereas that of table [\ref{m278}] is asymmetric. Hence, these two models
demonstrate that the reduction to three generations in itself 
does not require asymmetric boundary conditions. This seems to be
in contradiction  to the conclusion reached in \cite{fknr}. However,
I note that in [\ref{symmetric3gen}] the gauge group is broken
in two stages, whereas in \cite{fknr} the $SO(10)$ symmetry
remained unbroken. In particular, the breaking pattern
of the hidden gauge group $SO(16)\rightarrow SO(4)\times SO(12)$
in [\ref{symmetric3gen}] was a case not considered in the classification
of \cite{fknr}. This breaking pattern is, however, allowed
if the $SO(10)$ GUT symmetry is broken to $SO(6)\times SO(4)$,
as in [\ref{symmetric3gen}]. The conclusion is therefore that,
whereas the classification of ref. \cite{fknr} found that
the reduction to three generations necessitates the use
of an asymmetric shift in the scanned space of models,
more complicated symmetry breaking patterns allow for
the reduction also with symmetric actions. Hence, it is concluded
that the reduction to three generations in itself does not
necessitate the use of an asymmetric shift, in agreement with
the findings of ref. \cite{forste}. However, the distinction
between the symmetric versus asymmetric boundary conditions
is manifested in the models of table [\ref{symmetric3gen}]
versus table [\ref{m278}] at the more detailed level
of the spectrum and the phenomenological consequences. 
Hence, the world--sheet left--right symmetric model
of table [\ref{symmetric3gen}] produces $SU(3)$ color triplets
and does not give rise to untwisted--twisted--twisted
Standard Model fermion mass terms, whereas the asymmetric
model of table [\ref{m278}] does. As I discuss in the
following this distinction has a crucial implication also for
the issue of moduli fixing in these models. 

To translate the fermionic boundary conditions to twists and shifts in the 
bosonic formulation we bosonize the real fermionic degrees of freedom, 
$\{y,\omega\vert{\bar y},{\bar\omega}\}$. Defining, 
\beq
{\xi_i}={\sqrt{1\over2}}(y_i+i\omega_i)=e^{iX_i},
\eta_i=i{\sqrt{1\over2}}(y_i-i\omega_i)=ie^{-iX_i}
\label{xi}
\eeq
with similar definitions for the right movers $\{{\bar y},{\bar\omega}\}$
and $X^I(z,{\bar z})=X^I_L(z)+X^I_R({\bar z})$. 
With these definitions the world--sheet supercurrents in the bosonic 
and fermionic formulations are equivalent,
$$
T_F^{int}=\sum_{i}\chi_i{y_i}\omega_i=\sum_{i}\chi_i{\xi_i}\eta_i=
i\sum_{i}\chi_i\partial{X_i}.
$$
The momenta $P^I$ of the compactified scalars
in the bosonic formulation are identical with the $U(1)$ charges 
$Q(f)$ of the unbroken Cartan generators of the four dimensional
gauge group, eq. (\ref{qoff}).
The internal coordinates can be complexified by forming the 
combinations $(X=X_L+X_R)$
\beq
Z_k^\pm= (X^{2k-1}\pm i X^{2k}), ~~
\psi_k^\pm=(\chi^{2k-1}\pm i\chi^{2k}) ~~~(k=1,2,3)
\label{complexcombo}
\eeq
where the $Z_k^\pm$ are the complex coordinates of the six compactified
dimensions, now viewed as three complex planes, and $\psi^\pm_k$ are
the corresponding superpartners. 
\section{Moduli in free fermionic models}\label{moduliinffm}

The relevant moduli for our discussion here are the untwisted
K\"ahler and complex structure moduli of the six dimensional
compactified manifold. Additionally, the string vacua contain
the dilaton moduli whose VEV governs the strength of the
four dimensional interactions. The VEV of the dilaton moduli
is a continuous parameter from the point of view of the perturbative
heterotic string, and its stabilization requires some nonperturbative
dynamics, or some input from the underlying quantum M--theory,
which is not presently available. 
The problem of dilaton stabilization is therefore not addressed
in this paper, as the discussion here is confined to perturbative
heterotic string vacua. Since the moduli fields
correspond to scalar fields in the massless string spectrum,
the moduli space is determined by the set of boundary condition
basis vectors that define the string vacuum and encodes its
properties. The first step therefore is to identify the fields
in the fermionic models that correspond to the untwisted moduli.
The subsequent steps entail examining which moduli fields survive
successive GSO projections and consequently the residual moduli
space.  

The four dimensional fermionic heterotic string models are
described in terms of two dimensional conformal and superconformal
field theories of central charges $C_R=22$ and $C_L=9$, respectively.
In the fermionic formulation these are represented in terms
of world--sheet fermions. A convenient starting point
to formulate such a fermionic vacuum is a model
in which all the fermions are free. The free fermionic
formalism facilitates the solution of the conformal and
modular invariance constraints in terms of simple rules \cite{fff}.
Such a free fermionic model corresponds to a string vacuum
at a fixed point in the moduli space. Deformations from this
fixed point are then incorporated by including world-sheet
Thirring interactions among the world--sheet fermions,
that are compatible with the conformal and modular invariance
constraints. The coefficients of the allowed world--sheet Thirring
interactions correspond to the untwisted moduli fields.
For symmetric orbifold models, the exactly marginal operators
associated with the untwisted moduli fields take the general form
$\partial X^I{\bar\partial} X^J$, where $X^I$, $I=1,\cdots,6$, are
the coordinates of the six--torus $T^6$. Therefore, the untwisted
moduli fields in such models admit the geometrical interpretation of
background fields \cite{narain}, which appear as couplings of
the exactly marginal operators in the non--linear sigma model action,
which is the generating functional for string scattering amplitudes
\cite{cveticetal,cecottietal}.
The untwisted moduli scalar fields are the background fields that are 
compatible with the orbifold point group symmetry. 

It is noted that in the Frenkel--Kac--Segal construction
\cite{fks} of the Kac--Moody current algebra from chiral bosons,
the operator $i\partial X^I$ is a $U(1)$ generator of the 
Cartan sub--algebra. Therefore, in the fermionic
formalism the exactly marginal operators are given by Abelian
Thirring operators of the form $J_L^i(z){\bar J}_R^j({\bar z})$,
where $J_L^i(z)$, ${\bar J}_R^j({\bar z})$
are some left-- and right--moving $U(1)$ chiral currents
describe by world--sheet fermions. It has been shown that \cite{bagger}
Abelian Thirring interactions preserve conformal invariance,
and are therefore marginal operators. One can therefore
use the Abelian Thirring interactions to identify the
untwisted moduli in the free fermionic models. The
untwisted moduli correspond to the Abelian Thirring
interactions that are compatible with the GSO projections
induced by the boundary condition basis vectors, which
define the string models. 

The set of Abelian Thirring operators, and therefore of the 
untwisted moduli fields, is consequently restricted by progressive
boundary condition basis vector sets. The minimal basis set is
given by the basis that contains only the two vectors $\{1, S\}$. 
\begin{eqnarray}
1 &=& \{ \psi^{1,2}, \chi^{1,\ldots,6}, y^{1,\ldots,6}, \omega^{1,\ldots,6} |
\nonumber\\
& & \bar{y}^{1,\ldots,6}, \bar{\omega}^{1,\ldots,6}, \bar{\psi}^{1,\ldots,6},
\bar{\eta}^{1,2,3}, \bar{\phi}^{1,\ldots,8} \} ,\label{eq:1}\\
S &=& \{ \psi^{1,2}, \chi^{1,\ldots,6} \} .
\label{neq4basis}
\end{eqnarray}
This set generates a model with $N=4$ supersymmetry and $SO(44)$ 
right--moving gauge group
and correspond the a Narain model at an enhanced symmetry point. 
Accordingly, we can identify the six $\chi^I$ with the fermionic
superpartners of the six compactified bosonic coordinates.
Therefore, each pair $\{y^I, \omega^I\}$ is identified 
with the fermionized version of the corresponding left--moving
bosonic coordinate $X^I$, {\it i.e.} $i\partial X_L^I\sim y^I\omega^I$. 
The two--dimensional action for the for the Abelian Thirring
interactions is 
\beq
S~=~\int d^2z h_{ij}(X) J_L^i(z) {\bar J}_R^j({\bar z})~,
\label{2daction}
\eeq
where $J_L^i (i=1,\cdots ,6)$ are the chiral currents of the left--moving
$U(1)^6$ and ${\bar J}^j_R (j=1,\cdots,22)$, are the chiral currents
of the right--moving $U(1)^{22}$. The couplings $h_{ij}(X)$, as
functions of the space--times coordinates $X^\mu$, are four 
dimensional scalar fields that are identified with the scalar
components of the untwisted moduli fields. In the simplest model
with the two basis vectors of eq. (\ref{neq4basis})
the $6\times22$ fields $h_{ij}(X)$ in eq. (\ref{2daction})
are in one--to--one correspondence with the 21 and 15 components
of the background metric $G_{IJ}$ and antisymmetric tensor
$B_{IJ}$ $(I,J=1,\cdots,6)$, plus the $6\times 16$ Wilson lines
$A_{Ia}$. The $h_{ij}(X)$ fields therefore parameterize the
$SO(6,22)/SO(6)\times SO(22)$ coset--space of the toroidally
compactified manifold. The $h_{ij}$ untwisted moduli fields arise from
the Neveu--Schwarz sector, 
\beq
\vert \chi^I\rangle_L ~\otimes \vert{\bar\Phi}^{+J}{\bar\Phi}^{-J}\rangle_R~,
\label{untwisted22moduli}
\eeq
given in terms of the 22 complex right--handed world--sheet fermions
${\bar\Phi}^{+J}$ and their complex conjugates
${\bar\Phi}^{-J}$. The corresponding
marginal operators are given as
\beq
J^i_L(z){\bar J}^j_R({\bar z}) ~\equiv~~:y^i(z)\omega^i(z)(z):
:{\bar\Phi}^{+j}({\bar z}){\bar\Phi}^{-j}({\bar z}):~.
\label{marginal22operators}
\eeq
It is noted that the transformation properties of $\chi^i$, which appear
in the moduli (\ref{untwisted22moduli}), are the same as those of 
$y^i\omega^i$, which appear in the Abelian Thirring interactions
(\ref{marginal22operators}).

The next stages in the free fermionic model building consist of
adding consecutive boundary condition basis vectors. In the 
first instance we can add basis vectors that preserve the $N=4$
space--time supersymmetry, and with no periodic left--moving
fermions . Those that produce massless states may contain either
four, or eight, right--moving periodic fermions. At least
one basis vector with eight periodic right--moving fermions
is needed to produce space--time spinors under the GUT gauge
group. The free fermionic models correspond to $Z_2\times Z_2$
orbifolds and can then be classified
into models that produce spinorial representations from
one, two, or three of the twisted planes of the $Z_2\times Z_2$
orbifold \cite{fknr}. It turns out that the entire space of
models may be classified using a specific set of basis vectors,
and the varying cases are produced by the various choices of
the GSO projection coefficients. This covering basis contains
two basis vectors with eight non--overlapping right--moving periodic
fermions, and all left--moving world--sheet fermions are antiperiodic. 
Hence, with no loss of generality, these are the vector $\xi_1$
in eq. (\ref{vectorx1}) and the vector $\xi_2$
\beq
\xi_2=(0,\cdots,0\vert0,\cdots,0,
{\underbrace{1,\cdots,1}_{{\bar\phi^{1,\cdots,8}}}})~.
\label{vectorx2}
\eeq
where the notation introduced in section \ref{rffm} has been used.
{}We note that all the untwisted moduli and marginal operators in
eqs. (\ref{untwisted22moduli}) and (\ref{marginal22operators})
are invariant under the projections induced by (\ref{vectorx1})
and (\ref{vectorx2}).
The four dimensional right--moving gauge group is
now $SO(12)\times E_8\times E_8$. 

The next step in the construction is the inclusion
of the basis vectors $b_1$ and $b_2$ that correspond to the
action of the $Z_2\times Z_2$ orbifold twists. The assignment
of boundary condition in $b_1$ and $b_2$ may vary \cite{fknr},
and one specific choice is given by
\beqn
b_1 &=& ({\underbrace{1,\cdots\cdots\cdots,1}_
{{\psi^\mu},{\chi^{12}},y^{3,...,6},{\bar y}^{3,...,6}}},0,\cdots,0\vert
{\underbrace{1,\cdots,1}_{{\bar\psi}^{1,...,5},
{\bar\eta}^1}},0,\cdots,0) \\
b_2 &=& ({\underbrace{1,\cdots\cdots\cdots\cdots\cdots,1}_{
{\psi^\mu},{\chi^{34}},{y^{1,2}},{\omega^{5,6}},{{\bar y}^{1,2}}
{{\bar\omega}^{5,6}}}},0,\cdots,0\vert 
{\underbrace{1,\cdots,1}_{{{\bar\psi}^{1,...,5}},{\bar\eta}^2}},0,\cdots,0)
\label{b1b2}
\eeqn
In the notation of section \ref{rffm} we can write the
untwisted moduli fields, eq. (\ref{untwisted22moduli}) in the form
\beqn
& &\vert \chi^i\rangle_L~\otimes
\vert{\bar y}^{j}{\bar\omega}^{j}\rangle_R~,\\
& &\vert\chi^i\rangle_L~\otimes
\vert{\bar\Phi}^{+J}{\bar\Phi}^{-J}\rangle_R~,\\
\label{untwisted22moduli2}
\eeqn
and the (\ref{marginal22operators})
\beqn
& & J^i_L(z){\bar J}^j_R({\bar z}) ~\equiv~~ 
:y^i(z)\omega^i(z)::{\bar y}^j(z){\bar\omega}^j(z):~~~~(j=1,\cdots,6);\\
& & J^i_L(z){\bar J}^J_R({\bar z}) ~\equiv~~ 
:y^i(z)\omega^i(z):
:{\bar\Phi}^{+J}({\bar z}){\bar\Phi}^{-J}({\bar z}):~~~~(J=7,\cdots,22)~.
\label{marginal22operators2}
\eeqn
The effect of the additional basis vectors $b_1$ and $b_2$ is to make
some of the chiral currents, $J_L^i$ or ${\bar J}_R^j$,
antiperiodic. As a result some of the Abelian Thirring
interaction terms in eq. (\ref{2daction})
are not invariant when the world--sheet fermions are parallel
transported around the noncontractible loops of the world--sheet
torus. The corresponding moduli fields are projected from the 
massless spectrum by the GSO projections.

Under the basis vector $b_1$ as defined above the chiral currents
transform as: 
\beqn
& & J_L^{1,2}\rightarrow J_L^{1,2}~,~~~J_L^{3,4,5,6}\rightarrow
-J_L^{3,4,5,6}\\
& & {\bar J}_R^{1,2}\rightarrow
{\bar J}_R^{1,2}~,~~~{\bar J}_R^{3,4,5,6}\rightarrow 
-{\bar J}_R^{3,4,5,6}
\label{jrtojrb1}
\eeqn
and ${\bar J}_R^j(j=7,\cdots,22)$ are always periodic. Similarly,
under $b_2$ we have 
\beqn
& & J_L^{3,4}\rightarrow J_L^{3,4}~,~~~J_L^{1,2,5,6}\rightarrow
-J_L^{1,2,5,6}\\
& & {\bar J}_R^{1,2}\rightarrow
{\bar J}_R^{3,4}~,~~~{\bar J}_R^{1,2,5,6}\rightarrow 
-{\bar J}_R^{1,2,5,6}
\label{jrtojrb2}
\eeqn
As a consequence the only allowed Thirring interaction terms are
\beq
J_L^{1,2}{\bar J}_R^{1,2}, J_L^{3,4}{\bar J}^{3,4}, J_L^{5,6}
{\bar J}_R^{5,6}
\label{thirringz2xz2}
\eeq
Correspondingly there are three sets of untwisted moduli scalars
\beq
h_{ij}=\vert \chi^i\rangle_L \otimes 
\vert{\bar y}^j{\bar\omega}^j\rangle_R=
\cases{
(i,j=1,2)& $ $\cr
(i,j=3,4)& $ $\cr
(i,j=5,6)& $ $,\cr}
\label{hij}
\eeq
which parameterize the moduli space
\beq
{\cal M}=\left({{SO(2,2)}\over {SO(2)\times SO(2)}}\right)^3.
\eeq
We can define six complex moduli from the six real ones. For example,
on the first complex plane we can write,
\beqn
H^{(1)}_1 &=& {1\over \sqrt{2}}( h_{11}+ih_{21})={1\over \sqrt{2}}
\vert\chi^1+i\chi^2\rangle_L\otimes 
\vert{\bar y}^1{\bar\omega}^1\rangle_R\label{H11}\\
H^{(1)}_2 &=& {1\over \sqrt{2}}( h_{12}+ih_{22})={1\over \sqrt{2}}
\vert\chi^1+i\chi^2\rangle_L\otimes 
\vert{\bar y}^2{\bar\omega}^2\rangle_R
\label{H12}
\eeqn
These can be combined into three K\"ahler ($T_i$) structure and three
complex structure ($U_i$) moduli. For example, on the first complex
plane we define
\beqn
T_1 & = & {1\over \sqrt{2}}( H^{(1)}_1-iH^{(1)}_2)=
{1\over \sqrt{2}}(\chi^1+i\chi_2)\lv \otimes 
({\bar y}^1{\bar\omega}^1-i{\bar y}^2{\bar\omega}^2)\rv\label{t1u1}\\
U_1 & = & {1\over \sqrt{2}}( H^{(1)}_1+iH^{(1)}_2)=
{1\over \sqrt{2}}(\chi^1+i\chi_2)\lv \otimes 
({\bar y}^1{\bar\omega}^1+i{\bar y}^2{\bar\omega}^2)\rv
\label{u1t1}
\eeqn
and similarly for $T_{2,3}$ and $U_{2,3}$. These span
the coset moduli space
\beq
{\cal M}=\left[ {{SU(1,1)}\over{U(1)}}\otimes
{{SU(1,1)}\over{U(1)}}\right]^3~,
\eeq
which is the untwisted moduli space of the symmetric $Z_2\times Z_2$
orbifold. We can write the allowed
Thirring interaction terms, for the untwisted moduli of (\ref{hij}),
in terms of the complex coordinates $Z^\pm_k$.
For example, for the first set we have,
\beq
\sum_{i,j=1,2}h_{ij}{J}^i_L{\bar J}^j_R={1\over 2}\left(
T_1\partial Z^-_1{\bar\partial}Z^+_1+
{\bar T}_1\partial Z^+_1{\bar\partial}Z^-_1+
U_1\partial Z^-_1{\bar\partial}Z^-_1+
{\bar U}_1\partial Z^+_1{\bar\partial}Z^+\right)
\label{cthirring}
\eeq
where $T$ and $U$  are the complex fields defined in eqs. (\ref{t1u1})
and (\ref{u1t1}). The Thirring interactions for the other two complex
planes can be written similarly. 
The boundary condition basis vectors $b_1$ and $b_2$ translate into
twists of the complex planes \cite{foc}. Thus, $b_1$ leaves the first
complex plane invariant, and twists the second and third plane, {\it i.e.}
$Z_1\rightarrow Z_1$ and $Z_{2,3}\rightarrow -Z_{2,3}$, whereas $b_2$
leaves the second plane invariant and twists the first and third plane.
It is apparent that the Thirring terms in eq. (\ref{cthirring})
are invariant under the action of the $Z_2\times Z_2$ twists,
whereas all the mixed terms are projected out by the GSO projections.
{}From eq. (\ref{cthirring}) it is seen that the $T$ field is associated
the K\"ahler structure moduli, whereas the $U$ field is associated with a 
complex structure moduli. This agrees with the fact that the untwisted
sector of the $Z_2\times Z_2$ orbifold produces three complex and three 
K\"ahler structure moduli.

\section{Moduli in the three generation free fermionic models}\label{min3gm}

Next I turn to examine the untwisted moduli in the quasi--realistic
three generation free fermionic models. As discussed in section
\ref{rffm} the reduction to three generation is achieved by the 
inclusion of three or four additional boundary condition 
basis vectors, beyond the NAHE--set. This are typically denoted in
the literature as $b_i$ ($i=4,5,\cdots$) for basis vectors that
do not break the $SO(10)$ GUT symmetry, and by $\alpha,\beta,\gamma\cdots$,
for basis vectors that do. From the point of view of the untwisted moduli
beyond the NAHE--set models, the relevant boundary conditions are those
of the world--sheet fermions
$\{y,\omega\vert{\bar y},{\bar\omega}\}^{1,\cdots,6}$.

Left--right symmetric boundary conditions cannot eliminate any additional
moduli beyond those that exist in the NAHE--set model. This is a general
consequence of the requirement that the supercurrent (\ref{supercurrent})
is well defined under the parallel transport of the world--sheet fermions,
as well as the requirement of $N=1$ space--time supersymmetry.
It is instructive to recall that the moduli of the NAHE--based models
correspond to the  non--vanishing world--sheet Thirring interactions,
eq. (\ref{thirringz2xz2}), and 
are those of the $Z_2\times Z_2$ orbifold, eq. (\ref{cthirring}).
An example of a left--right three generation free fermionic model is given
in table [\ref{symmetric3gen}]. Rewriting the boundary conditions for the
internal world--sheet fermions in the notation of table
[\ref{symmetricthirring}] 
\begin{eqnarray}
     &\begin{tabular}{c|cccc|cccc|cccc}
 ~&   $y^1{\omega}^1$
  &   ${\bar y}^1{\bar\omega}^1$
  &   $y^2{\omega}^2$
  &   ${\bar y}^2{\bar\omega}^2$
  &   $y^3{\omega}^3$
  &   ${\bar y}^3{\bar\omega}^3$
  &   $y^4 {\omega}^4$
  &   ${\bar y}^4{\bar\omega}^4$
  &   $y^5{\omega}^5$
  &   ${\bar y}^5{\bar\omega}^5$
  &   $y^6{\omega}^6$
  &   ${\bar y}^6{\bar\omega}^6$ \\
\hline
\hline
$\alpha$ & 00 & 00 & 00 & 00  & 10 & 10 & 01 & 01  & 01 & 01 & 10 & 10
\\
$\beta$  & 01 & 01 & 10 & 10  & 00 & 00 & 00 & 00  & 01 & 01 & 10 & 10
\\
$\gamma$ & 10 & 10 & 01 & 01  & 10 & 10 & 10 & 10  & 00 & 00 & 00 & 00
\\\nonumber
\end{tabular}\\
\label{symmetricthirring}
\end{eqnarray}
it is evident that all the Thirring interaction terms in eq.
(\ref{thirringz2xz2}) are invariant under the transformations in
table [\ref{symmetricthirring}]. As discussed in \cite{lny}
and section \ref{moduliinffm} there is a one--to--one correspondence
between the Thirring terms, eq. (\ref{thirringz2xz2}) and the moduli fields, 
eq. (\ref{hij}). Therefore, it is sufficient to examine either
the moduli or the Thirring terms, and this relation guarantees the
existence, or exclusion, of the corresponding Thirring terms/moduli fields.

Left--right symmetric boundary conditions preserve the untwisted
moduli space of the $Z_2\times Z_2$
orbifold. To see that this is indeed a general result, and a consequence of 
world--sheet, and space--time, supersymmetry, let us recall that the
world--sheet supercurrent eq. (\ref{supercurrent}) restricts that
each triplet $\chi_iy_i\omega_i$ must transform as $\psi^\mu\partial X_\mu$.
The boundary conditions of each triplets are therefore restricted
to be
\beqn
b(\psi^\mu)& = & 1\rightarrow b(\chi_i,y_i,\omega_i)=
			(1,0,0);~(0,1,0);~(0,0,1);~(1,1,1)\cr
b(\psi^\mu)& = & 0\rightarrow b(\chi_i,y_i,\omega_i)=
			(0,1,1);~(1,0,1);~(1,1,0);~(0,0,0)
\label{restonbcfsc}
\eeqn 
The requirement of $N=1$ supersymmetry further restricts that 
\beq
b(\chi_{2k-1})~=~b(\chi_{2k})~~~k=1,2,3
\eeq
Symmetric boundary conditions means that the boundary condition
of the left--moving $b\{y_i,\omega_i\}$ are identical to their
corresponding right--moving fields $b\{{\bar y}_i,{\bar\omega}_i\}$.
In terms of the internal conformal field theory this means that 
symmetric boundary conditions correspond to the case in which
all the internal left--moving world--sheet fermions from the set
$\{y, \omega\}^{1,\cdots,6}$ are paired with 12 right--moving world--sheet
real fermions $\{{\bar y}, {\bar\omega}\}^{1,\cdots,6}$, to form 
12 Ising model operators \cite{kln, fm}. 
Enumerating all the possible boundary conditions for the internal 
fermions, subject to these restrictions we have
\beqn
& b(\psi^\mu)=1 ~\&~ b(\chi_{2k-1}=\chi_{2k})=1\rightarrow
 b(\{y,\omega\}_{2k-1},\{y,\omega\}_{2k})=\cases{
						0~ 0~ 0~ 0 &\ $ $\cr
						0~ 0~ 1~ 1 &\ $ $\cr
						1~ 1~ 0~ 0 &\ $ $\cr
						1~ 1~ 1~ 1 &\ $ $\cr}\cr
\nonumber\\
\nonumber\\
& b(\psi^\mu)=1 ~\&~ b(\chi_{2k-1}=\chi_{2k})=0\rightarrow
 b(\{y,\omega\}_{2k-1},\{y,\omega\}_{2k})=\cases{
						1~ 0~ 1~ 0 &\ $ $\cr
						1~ 0~ 0~ 1 &\ $ $\cr
						0~ 1~ 1~ 0 &\ $ $\cr
						0~ 1~ 0~ 1 &\ $ $\cr}\cr
\nonumber\\
\nonumber\\
& b(\psi^\mu)=0 ~\&~ b(\chi_{2k-1}=\chi_{2k})=1\rightarrow
 b(\{y,\omega\}_{2k-1},\{y,\omega\}_{2k})=\cases{
						1~ 0~ 1~ 0 &\ $ $\cr
						1~ 0~ 0~ 1 &\ $ $\cr
						0~ 1~ 1~ 0 &\ $ $\cr
						0~ 1~ 0~ 1 &\ $ $\cr}\cr
\nonumber\\
\nonumber\\
& b(\psi^\mu)=0 ~\&~ b(\chi_{2k-1}=\chi_{2k})=0\rightarrow
 b(\{y,\omega\}_{2k-1},\{y,\omega\}_{2k})=\cases{
						0~ 0~ 0~ 0 &\ $ $\cr
						0~ 0~ 1~ 1 &\ $ $\cr
						1~ 1~ 0~ 0 &\ $ $\cr
						1~ 1~ 1~ 1 &\ $ $\cr}
\label{symmetricpossibilities}
\eeqn
{\it i.e.} there are in total four different possibilities
for the transformation of the pairs $\{y_i,\omega_i\}$
\beq
b(\{y_i,\omega_i\})~=~ (0,0);~ (1,0);~ (0,1);~ (1,1)
\eeq
Using the relations (\ref{xi}) these translate to transformations 
of the corresponding bosonic coordinates, {\it i.e.}
\beq
  \begin{tabular}{|cc|c|c|c|}
\hline
  &   $y,{\omega}$
  &   $X_L$
  &   $X_R$
  &   $X=X_L+X_R$ \\
\hline
\hline
& 0 0 & $X_L+\pi$  & $X_R+\pi$  & $X+2\pi$  \\
& 1 0 & $-X_L$     & $-X_R$     & $-X$      \\
& 0 1 & $-X_L+\pi$ & $-X_R+\pi$ & $-X+2\pi$ \\
& 1 1 & $X_L$      & $X_R$      & $X$       \\
\hline
\end{tabular}
\eeq
In terms of the complexified coordinates $Z_k$, eq. (\ref{complexcombo})
these translate into 
\beq
Z_k\rightarrow \pm Z_k+\delta_12\pi+\delta_2i2\pi,
\label{complextransform}
\eeq
where $\delta_{1,2}=0;1$ signify possible shifts in the real and
complex directions. 
Examining the complexified Thirring terms, eq. (\ref{cthirring}),
it is noted that these always involve pairs of complex coordinates
that transform with the same sign, and hence are always invariant
under symmetric boundary conditions. 

Relaxing the condition of $N=1$ space--time supersymmetry means
that we may have 
\beq
b(\chi_{2k-1})~\ne~b(\chi_{2k})
\eeq
for some $k$. For example, for $k=1$ we may have
\beq
  \begin{tabular}{cccc}
    $y_1$  & $\omega_1$  & $y_2$  & $\omega_2$ \\
     1     &    1        &   1    &   0        \\
\end{tabular}
\eeq
In this case the moduli fields $h_{11}$ and $h_{22}$ are retained,
whereas  $h_{12}$ and $h_{21}$ are projected out. Hence, if we break
$N=1$ space--time supersymmetry by the assignment of boundary conditions
we may project six additional untwisted moduli, and six will remain.

Next, I turn to examine the presence of untwisted moduli 
in models with asymmetric boundary conditions. The assignment
of asymmetric boundary conditions entails that a left--moving
real fermion from the set $\{y,\omega\}$ is paired with another
left--moving real fermion from this set, and with which it has
identical boundary conditions in all basis vectors. For every
such pair of left--moving fermions, there is a corresponding
pair of right--moving fermions. These combinations therefore 
give rise to a global $U(1)_L$ symmetry, and a corresponding
gauged $U(1)_R$ symmetry. These complex pairings therefore allow
the assignment of asymmetric boundary conditions, with important
phenomenological consequences \cite{cslm,ps,topbottomyuk,pairings}.
The simplest case is that of table [\ref{model2}] which involves
a single such complexified fermion. 
%
%
%
%
\begin{eqnarray}
 &\begin{tabular}{c|c|ccc|c|ccc|c}
 ~ & $\psi^\mu$ & $\chi^{12}$ & $\chi^{34}$ & $\chi^{56}$ &
        $\bar{\psi}^{1,...,5} $ &
        $\bar{\eta}^1 $&
        $\bar{\eta}^2 $&
        $\bar{\eta}^3 $&
        $\bar{\phi}^{1,...,8} $\\
\hline
\hline
  ${\alpha}$     &  1 & 1&0&0 & 1~1~1~0~0 & 1 & 0 & 1 & 1~1~1~1~0~0~0~0 \\
  ${\beta}$      &  1 & 0&1&0 & 1~1~1~0~0 & 0 & 1 & 1 & 1~1~1~1~0~0~0~0 \\
  ${\gamma}$     &  1 & 0&0&1 &
		${1\over2}$~${1\over2}$~${1\over2}$~${1\over2}$~${1\over2}$
	      & ${1\over2}$ & ${1\over2}$ & ${1\over2}$ 
	      & ${1\over2}$~0~1~1~${1\over2}$~${1\over2}$~${1\over2}$~0 \\
\end{tabular}
   \nonumber\\
   ~  &  ~ \nonumber\\
   ~  &  ~ \nonumber\\
     &\begin{tabular}{c|c|c|c}
 ~&   $y^3{\bar y}^3$
      $y^4{\bar y}^4$
      $y^5{\bar y}^5$
      ${y}^6{\bar y}^6$
  &   $y^1{\bar y}^1$
      $y^2{\bar y}^2$
      $\omega^5{\bar\omega}^5$
      ${\omega}^6{\bar\omega}^6$
  &   $\omega^2{\omega}^3$
      $\omega^1{\bar\omega}^2$
      $\omega^4{\bar\omega}^4$
      ${\bar\omega}^2{\bar\omega}^3$ \\
\hline
\hline
$\alpha$ & 1 ~~~ 0 ~~~ 0 ~~~ 1  & 0 ~~~ 0 ~~~ 1 ~~~ 0  & 0 ~~~ 0 ~~~ 1 ~~~ 1
\\
$\beta$  & 0 ~~~ 0 ~~~ 0 ~~~ 1  & 0 ~~~ 1 ~~~ 1 ~~~ 0  & 0 ~~~ 1 ~~~ 0 ~~~ 1
\\
$\gamma$ & 1 ~~~ 1 ~~~ 0 ~~~ 0  & 1 ~~~ 1 ~~~ 0 ~~~ 0  & 0 ~~~ 0 ~~~ 0 ~~~ 1
\\
\end{tabular}
\label{model2}
\end{eqnarray}
With some choice of GSO projection coefficients.
The model of table [\ref{model2}] was published in \cite{cslm},
and the entire spectrum is given there.
In this model it is easily seen
that the moduli fields,
\beq
h_{11},~h_{21},~h_{34},~h_{44},~
h_{55},~h_{56},~h_{65},~h_{66},~
\label{retmodmod2}
\eeq
and their corresponding Thirring terms, are retained in the spectrum,
whereas the moduli fields, 
\beq
h_{12},~h_{22},~h_{33},~h_{43},~
\label{promodmod2}
\eeq
are projected out, and the corresponding Thirring terms are 
not invariant under the transformations.
Using the bosonic coordinates eq. (\ref{complexcombo})
we can translate the transformations of the internal fermions
to the corresponding action on bosonic variables.
In terms of
\beq
Z_i=X_{2i-1}+iX_{2i}=X_{2i-1}^L+X_{2i-1}^R+i(X_{2i}^L+X_{2i}^R)~~~(i=1,2,3),
\label{comcom}
\eeq
we have
\beq
     \begin{tabular}{c|c|c|c}
      &   $Z_1$     &      $Z_2$        &  $Z_3$ \\
\hline
\hline
$\alpha$ & 
		$X_1+i(X_2^L-X_2^R)+2\pi+i2\pi$ & 
		$-X_3^L+X_3^R-iX_4+i2\pi$ & 
		$-Z_3+2\pi$ \\
$\beta$  &
		$-X_1+i(-X_2^L+X_2^R)+2\pi$ & 
		$X_3^L-X_3^R-iX_4+2\pi+i2\pi$ & 
		$-Z_3+2\pi$ \\
$\gamma$ & 
		$-X_1+i(-X_2^L+X_2^R)$ & 
		$-X_3^L+X_3^R-iX_4$ & 
		$Z_3+2\pi+i2\pi$ 
\end{tabular}
\label{model2complexbc}
\eeq
{}From table [\ref{model2complexbc}] we see that the complex structure
of two of the complex planes is broken, whereas the third is retained.
Furthermore, the transformations with respect to two of the internal
bosonic coordinates are asymmetric between the left-- and the right--moving
part, and are therefore not geometrical. This entails that the corresponding
moduli must be projected out, and the coordinates are frozen at fixed
radii. Indeed, using the identity (\ref{cthirring}), 
and the definitions of the complex moduli fields in eq.
(\ref{H11}--\ref{u1t1}) we relate
the projection of the moduli fields $h_{ij}$ to constraints
on the complex and K\"ahler moduli of the $Z_2\times Z_2$ orbifold,
$U_i$ and $T_i$ $(i=1,2,3)$. Thus, in the case of the model of
table [\ref{model2}], the projection of the
$h_{11},~h_{22},~h_{33},~h_{43}$ moduli fields translate to
the conditions
\beq
T_1=U_1~~~,~~~T_2=-U_2~~~ {\rm whereas}~~~ T_3;~=U_3~~~{\rm are~unconstrained}
\label{model2complexmod}
\eeq

The next example is a model with two complexified left-- and right--moving
internal fermions from the set $\{y,\omega\vert{\bar y},{\bar\omega}\}$.
Table [\ref{dicomplexfer}] provides an example of such a model.
\beqn
 &\begin{tabular}{c|c|ccc|c|ccc|c}
 ~ & $\psi^\mu$ & $\chi^{12}$ & $\chi^{34}$ & $\chi^{56}$ &
        $\bar{\psi}^{1,...,5} $ &
        $\bar{\eta}^1 $&
        $\bar{\eta}^2 $&
        $\bar{\eta}^3 $&
        $\bar{\phi}^{1,...,8} $ \\
\hline
\hline
  $\alpha$    &  0 & 0&0&0 & 1~1~1~0~0 & 0 & 0 & 1 & 1~1~1~1~0~0~0~0 \\
  $\beta$     &  0 & 0&0&0 & 1~1~1~0~0 & 0 & 0 & 1 & 1~1~1~1~0~0~0~0 \\
  ${\gamma}$  &  0 & 0&0&0 &
		${1\over2}$~${1\over2}$~${1\over2}$~${1\over2}$~${1\over2}$
	      & ${1\over2}$ & ${1\over2}$ & ${1\over2}$ &
                ${1\over2}$~${1\over2}$~0~0~0~${1\over2}$~${1\over2}$~1 \\
\end{tabular}
   \nonumber\\
   ~  &  ~ \nonumber\\
   ~  &  ~ \nonumber\\
     &\begin{tabular}{c|c|c|c}
 ~&   $y^3{y}^6$
      $y^4{\bar y}^4$
      $y^5{\bar y}^5$
      ${\bar y}^3{\bar y}^6$
  &   $y^1{\bar\omega}^5$
      $y^2{\bar y}^2$
      $\omega^6{\bar\omega}^6$
      ${\bar y}^1{\bar\omega}^5$
  &   $\omega^1{\bar\omega}^1$
      $\omega^2{\bar\omega}^2$
      $\omega^3{\bar\omega}^3$
      ${\omega}^4{\bar\omega}^4$ \\
\hline
\hline
$\alpha$ & 1 ~~~ 0 ~~~ 0 ~~~ 0  & 0 ~~~ 0 ~~~ 1 ~~~ 1  & 0 ~~~ 0 ~~~ 1 ~~~ 0 \\
$\beta$  & 0 ~~~ 0 ~~~ 1 ~~~ 1  & 1 ~~~ 0 ~~~ 0 ~~~ 0  & 1 ~~~ 0 ~~~ 0 ~~~ 0 \\
$\gamma$ & 0 ~~~ 1 ~~~ 0 ~~~ 1  & 0 ~~~ 0 ~~~ 0 ~~~ 1  & 0 ~~~ 0 ~~~ 0 ~~~ 1 \\
\end{tabular}
\label{dicomplexfer}
\eeqn
With some choice of generalized GSO coefficients.
The model of table [\ref{dicomplexfer}] produces three generations from
the twisted sector $b_1$, $b_2$ and $b_3$, and a standard--like observable
gauge group. It produces Electroweak super--Higgs doublets from the first two
untwisted planes, and a color triplet from the third. 
In this model it is easily seen that the moduli fields,
\beq
h_{12},~h_{22},~h_{34},~h_{44},~
\label{retmodmoddiq}
\eeq
and their corresponding Thirring terms, are retained in the spectrum,
whereas the moduli fields, 
\beq
h_{11},~h_{22},~h_{33},~h_{43},~
h_{55},~h_{56},~h_{65},~h_{66},~
\label{promodmoddiq}
\eeq
are projected out, and the corresponding Thirring terms are 
not invariant under the transformations. In terms of the complex
bosonic coordinates (\ref{comcom}) we have
on the first two complex planes
\beq
     \begin{tabular}{c|c|c}
      &   $Z_1$     &      $Z_2$     \\
\hline
\hline
$\alpha$ & 
		$(X^L_1-X^R_1)+iX_2+(1+2i)\pi$ & 
		$(X^L_3-X^R_3)+iX_4+(1+2i)\pi$  \\
$\beta$  &
		$(X^L_1-X^R_1)+iX_2+(1+2i)\pi$ & 
		$(X^L_3-X^R_3)+iX_4+(1+2i)\pi$ \\
$\gamma$ & 
		$(X^L_1-X^R_1)+iX_2+(1+2i)\pi$ & 
		$(X^L_3-X^R_3)+iX_4+\pi$ 
\end{tabular}
\label{model2comfercombc12}
\eeq
and on the third
\beq
     \begin{tabular}{c|c}
     &  $Z_3$ \\
\hline
\hline
$\alpha$ & 
		$(X^L_5-X^R_5)+i(X^L_6-X^R_6)+(2+i)\pi$  \\
$\beta$  &
		$(X^L_5-X^R_5)+i(X^L_6-X^R_6)+i\pi$  \\
$\gamma$ &
		$(X^L_5-X^R_5)+i(X^L_6-X^R_6)+(2+i)\pi$
\end{tabular}
\nonumber 
\eeq
the projection of the
moduli fields in eq. (\ref{promodmoddiq}) translate in this case to
the conditions
\beq
T_1=-U_1~~~,~~~T_2=-U_2~~~ {\rm whereas}~~~ T_3;~=U_3~=0
\label{model2comfercommod}
\eeq
It is seen here that the third complex plane is completely fixed,
whereas on the first and second planes $X_2$ and $X_4$ retain
their geometrical character, and  $X_1$ and $X_3$ do not.

{}From the two examples above, and eq. (\ref{cthirring})
we can draw the general constraints on the complex and K\"ahler moduli,
which can be written for $k=1,2,3$ as,
\beqn
U_k & = & h_{2k-1~2k-1}+h_{2k~2k}+i(h_{2k~2k-1}-h_{2k-1~2k}) \label{uk}\\
T_k & = & h_{2k-1~2k-1}-h_{2k~2k}+i(h_{2k~2k-1}+h_{2k-1~2k}). \label{tk}
\eeqn
{}From the identity eq. (\ref{cthirring}), the vanishing of certain
real Thirring terms on the left--hand side, translate to the conditions
\beqn
T_k+U_k&=&0~~\Leftrightarrow~~h_{2k-1~2k-1} ~\&~ h_{2k~2k-1}~~
{\rm are~projected~out}.\label{tkplusuk}\\
T_k-U_k&=&0~~\Leftrightarrow~~~h_{2k~2k} ~~~~~\&~ h_{2k-1~2k}~~
{\rm are~projected~out}.\label{tkminusuk}
\eeqn
Hence, if both $T_k+U_k=0$ and $T_k-U_k=0$ hold, then both the
K\"ahler and complex structure moduli of the $k^{th}$ plane are projected
out, and the radii and angles of the corresponding internal torus
are frozen. In this case there is no extended geometry in the 
effective low energy field theory. This situation is similar
to the manner in which gauge symmetries are broken in string theory
by the GSO projections. Namely, the gauge symmetry is not realized
in the effective low energy field theory, but exists as an organizing
principle at the string theory level. That is parts of the string
spectrum obeys the symmetry, but the entire string spectrum does not. 

The above two examples
demonstrates the interesting possibility of correlating between
the number of complexified fermions and the surviving untwisted
moduli, which would suggest that for every complex fermion,
four additional untwisted moduli are projected out. However,
in the following I show that this is not necessarily the case,
and the situation is more intricate. 

The retention or projection of untwisted moduli in the case with three 
complex fermions depends on the choice of pairings of the left--moving
real fermions from the set $\{ y, \omega\}^{1,\cdots,6}$. The distinction
between the different choices of pairings, and some phenomenological
consequences, was briefly discussed in ref. \cite{pairings}.
To demonstrate the effect on the untwisted moduli I consider the
two models in tables [\ref{m278}] \cite{eu} and [\ref{fnymodel}] \cite{fny}.

\begin{eqnarray}
 &\begin{tabular}{c|c|ccc|c|ccc|c}
 ~ & $\psi^\mu$ & $\chi^{12}$ & $\chi^{34}$ & $\chi^{56}$ &
        $\bar{\psi}^{1,...,5} $ &
        $\bar{\eta}^1 $&
        $\bar{\eta}^2 $&
        $\bar{\eta}^3 $&
        $\bar{\phi}^{1,...,8} $\\
\hline
\hline
  ${b_4}$     &  1 & 1&0&0 & 1~1~1~1~1 & 1 & 0 & 0 & 0~0~0~0~0~0~0~0 \\
  ${\alpha}$   &  1 & 0&0&1 & 1~1~1~0~0 & 1 & 0 & 1 & 1~1~1~1~0~0~0~0 \\
  ${\beta}$  &  1 & 0&1&0 &
		${1\over2}$~${1\over2}$~${1\over2}$~${1\over2}$~${1\over2}$
	      & ${1\over2}$ & ${1\over2}$ & ${1\over2}$ 
	      & ${1\over2}$~0~1~1~${1\over2}$~${1\over2}$~${1\over2}$~1 \\
\end{tabular}
   \nonumber\\
   ~  &  ~ \nonumber\\
   ~  &  ~ \nonumber\\
     &\begin{tabular}{c|c|c|c}
 ~&   $y^3{y}^6$
      $y^4{\bar y}^4$
      $y^5{\bar y}^5$
      ${\bar y}^3{\bar y}^6$
  &   $y^1{\omega}^6$
      $y^2{\bar y}^2$
      $\omega^5{\bar\omega}^5$
      ${\bar y}^1{\bar\omega}^6$
  &   $\omega^1{\omega}^3$
      $\omega^2{\bar\omega}^2$
      $\omega^4{\bar\omega}^4$
      ${\bar\omega}^1{\bar\omega}^3$ \\
\hline
\hline
$\alpha$ & 1 ~~~ 0 ~~~ 0 ~~~ 1  & 0 ~~~ 0 ~~~ 1 ~~~ 0  & 0 ~~~ 0 ~~~ 1 ~~~ 0
\\
$\beta$  & 0 ~~~ 0 ~~~ 0 ~~~ 1  & 0 ~~~ 1 ~~~ 0 ~~~ 1  & 1 ~~~ 0 ~~~ 1 ~~~ 0
\\
$\gamma$ & 0 ~~~ 0 ~~~ 1 ~~~ 1  & 1 ~~~ 0 ~~~ 0 ~~~ 1  & 0 ~~~ 1 ~~~ 0 ~~~ 0
\\
\end{tabular}
\label{fnymodel}
\end{eqnarray}
The choice of generalized GSO projection coefficients of the model
of table [\ref{fnymodel}], as well as the 
complete mass spectrum and its charges under the four dimensional
gauge group are given in ref. \cite{fny}.

One of the important constraints in the construction
of the free fermionic models is the requirement that the 
supercurrent, eq. (\ref{supercurrent}), is well defined.
In the models that utilize
only periodic and anti--periodic boundary conditions
for the left--moving sector, the eighteen left--moving fermions
are divided into six triplets in the adjoint representation
of the automorphism group $SU(2)^6$.
These triplets are denoted by $\{\chi_i,y_i,\omega_i\}$
$i=1,\cdots,6$, and their boundary conditions are constrained
as given in eq. (\ref{restonbcfsc}).
In the type of models that are considered here a pair of real
fermions which are combined to form a complex fermion
or an Ising model operator must have the identical boundary
conditions in all sectors. In practice it is sufficient
to require that a pair of such real fermions
have the same boundary conditions in all the boundary
basis vectors which span a given model. In the model of table 
[\ref{m278}] the pairings are:
\beqn
&& \{(y^3y^6,y^4{\bar y}^4,y^5{\bar y}^5,{\bar y^3}{\bar y}^6), \nonumber\\
&& (y^1\omega^6,y^2{\bar y}^2,\omega^5{\bar\omega}^5,
			{\bar y}^1{\bar\omega}^6),\nonumber\\
&& (\omega^2\omega^4,\omega^1{\bar\omega}^1,\omega^3{\bar\omega}^3,
				{\bar\omega^2}{\bar\omega}^4)\},
\label{m278pairing}
\eeqn
where the notation emphasizes the original division of the
world--sheet fermions by the NAHE--set \cite{nahe,pairings}.
Note that with this pairing the complexified left-moving pairs
mix between the six $SU(2)$ triplets of the
left--moving automorphism group. That is the boundary condition
of $y^3y^6$ fixes the boundary condition of $y^1\omega^5$.
In this model we find that under the $\alpha$ projection
\beq
J^{1,\cdots,6}_L\rightarrow -J^{1,\cdots,6}_L,
{\bar J}^{1,\cdots,6}_R\rightarrow  {\bar J}^{1,\cdots,6}_R~.
\eeq
As a result all of the Thirring terms, and hence all the untwisted moduli
are projected out in this model.
In terms of the complex
bosonic coordinates (\ref{comcom}) the fermionic boundary conditions
on the first two complex planes translate to
\beq
     \begin{tabular}{c|c|c}
      &   $Z_1$     &      $Z_2$     \\
\hline
\hline
$\alpha$ & 
		$(X^L_1-X^R_1)+i(X^L_2-X^R_2)+(1+2i)\pi$ & 
		$(X^L_3-X^R_3)+i(X^L_4-X^R_4)+(1+2i)\pi$  \\
$\beta$  &
		$(X^L_1-X^R_1)+i(X^L_2-X^R_2)+(1+2i)\pi$ & 
		$(X^L_3-X^R_3)+i(X^L_4-X^R_4)+(1+2i)\pi$ \\
$\gamma$ & 
		$(X^L_1-X^R_1)+i(X^L_2-X^R_2)+\pi$ & 
		$(X^L_3-X^R_3)+i(X^L_4-X^R_4)+\pi$ 
\end{tabular}
\label{m278comfercombc12}
\eeq
and on the third
\beq
     \begin{tabular}{c|c}
     &  $Z_3$ \\
\hline
\hline
$\alpha$ & 
		$(X^L_5-X^R_5)+i(X^L_6-X^R_6)+(2+i)\pi$  \\
$\beta$  &
		$(X^L_5-X^R_5)+i(X^L_6-X^R_6)+i\pi$  \\
$\gamma$ &
		$(X^L_5-X^R_5)+i(X^L_6-X^R_6)+(2+i)\pi$
\end{tabular}
\nonumber 
\eeq
Hence, the boundary conditions in this case correspond to asymmetric
action on all six real internal coordinates. In terms
of the K\"ahler and complex structure fields we have that the
projection of the 12 real $h_{ij}$ translate to
\beq
T_k+U_k=0~~{\rm and}~~ T_k-U_k=0~~{\rm for}~~k=1,2,3~.
\eeq
Therefore, $$T_k=U_k=0.$$
and, and all the untwisted geometrical moduli are projected out
in this model.  

On the other hand the pairings in the
model of table [\ref{fnymodel}] are: 
\beqn
&& \{(y^3y^6,y^4{\bar y}^4,y^5{\bar y}^5,{\bar y^3}{\bar y}^6),\nonumber\\
&& (y^1\omega^6,y^2{\bar y}^2,\omega^5{\bar\omega}^5,
			{\bar y}^1{\bar\omega}^6),\nonumber\\
&& (\omega^1\omega^3,\omega^2{\bar\omega}^2,\omega^4{\bar\omega}^4,
				{\bar\omega^1}{\bar\omega}^3)\}
\label{fnypairing}
\eeqn
Note that with this pairing the complexified left-moving pairs
in Eqs. (\ref{fnypairing})
mix between the first, third and sixth $SU(2)$ triplets of the
left--moving automorphism group.
In this model it is easily seen that the moduli fields,
\beq
h_{11},~h_{12},~h_{21},~h_{22},~
h_{34},~h_{44},~h_{55},~h_{65}
\label{retmodmodfny}
\eeq
and their corresponding Thirring terms, are retained in the spectrum,
whereas the moduli fields, 
\beq
h_{33},~h_{43},~h_{56},~h_{66}
\label{promodmodfny}
\eeq
are projected out, and the corresponding Thirring terms are 
not invariant under the transformations.
In terms of the bosonized
variables the boundary conditions translate to
\beq
     \begin{tabular}{c|c|c}
      &   $Z_1$     &      $Z_2$     \\
\hline
\hline
$\alpha$ &
                $Z_1+(2+i2)\pi$ &  
                $-Z_2+i2\pi$  \\
$\beta$  &
                $-Z_1+\pi$ &
                $-Z_2+(2+i)\pi$ \\
$\gamma$ &   
                $-Z_1+i2\pi$ &
                $(X^L_3-X^R_3)+iX_4+(1+i2)\pi$
\end{tabular}
\label{fnycomfercombc12}
\eeq
and on the third
\beq
     \begin{tabular}{c|c}
     &  $Z_3$ \\
\hline
\hline
$\alpha$ &
                $-Z_3+2\pi$  \\
$\beta$  &
                $Z_3+(2+i)\pi$  \\
$\gamma$ &
                $-X_5+i(-X^L_6+X^R_6)+i\pi$
\end{tabular}
\nonumber 
\eeq
Hence, in this case the asymmetric fermionic boundary conditions
translate to asymmetric action only on two of the six real bosonic
coordinates, whereas the other four retain their geometrical
interpretation. In terms of the K\"ahler and complex structure 
fields the projection of the 4 real $h_{ij}$ fields in eq.
(\ref{promodmodfny}) translate to
\beq
T_1;~U_1~~~{\rm are~unconstrained},~~~T_2=-U_2~~~,~~~T_3=U_3~~~
\label{fnycomplexmod}
\eeq

Furthermore, with the choice of pairings in eqs. (\ref{fnypairing})
we can construct a model in
which all the untwisted moduli are retained in the spectrum.
An example of such a model is given in table [\ref{fnyvariant}].
\begin{eqnarray}
 &\begin{tabular}{c|c|ccc|c|ccc|c}
 ~ & $\psi^\mu$ & $\chi^{12}$ & $\chi^{34}$ & $\chi^{56}$ &
        $\bar{\psi}^{1,...,5} $ &
        $\bar{\eta}^1 $&
        $\bar{\eta}^2 $&
        $\bar{\eta}^3 $&
        $\bar{\phi}^{1,...,8} $\\
\hline
\hline
  ${\alpha}$     &  0 & 0&0&0 & 1~1~1~0~0 & 0 & 0 & 0 & 1~1~1~1~0~0~0~0 \\
  ${\beta}$      &  0 & 0&0&0 & 1~1~1~0~0 & 0 & 0 & 0 & 1~0~0~0~0~0~0~1 \\
  ${\gamma}$     &  0 & 0&0&0 &
		${1\over2}$~${1\over2}$~${1\over2}$~${1\over2}$~${1\over2}$
	      & ${1\over2}$ & ${1\over2}$ & ${1\over2}$ 
	      & ${1\over2}$~0~1~1~${1\over2}$~${1\over2}$~${1\over2}$~1 \\
\end{tabular}
   \nonumber\\
   ~  &  ~ \nonumber\\
   ~  &  ~ \nonumber\\
     &\begin{tabular}{c|c|c|c}
 ~&   $y^3{y}^6$
      $y^4{\bar y}^4$
      $y^5{\bar y}^5$
      ${\bar y}^3{\bar y}^6$
  &   $y^1{\omega}^6$
      $y^2{\bar y}^2$
      $\omega^5{\bar\omega}^5$
      ${\bar y}^1{\bar\omega}^6$
  &   $\omega^1{\omega}^3$
      $\omega^2{\bar\omega}^2$
      $\omega^4{\bar\omega}^4$
      ${\bar\omega}^1{\bar\omega}^3$ \\
\hline
\hline
$\alpha$ & 1 ~~~ 0 ~~~ 0 ~~~ 0  & 1 ~~~ 0 ~~~ 0 ~~~ 0  & 1 ~~~ 0 ~~~ 0 ~~~ 0
\\
$\beta$  & 0 ~~~ 1 ~~~ 0 ~~~ 1  & 0 ~~~ 0 ~~~ 0 ~~~ 1  & 0 ~~~ 0 ~~~ 1 ~~~ 1
\\
$\gamma$ & 0 ~~~ 0 ~~~ 1 ~~~ 1  & 0 ~~~ 0 ~~~ 1 ~~~ 1  & 0 ~~~ 0 ~~~ 0 ~~~ 1
\\
\end{tabular}
\label{fnyvariant}
\end{eqnarray}
with a choice of generalized GSO projection coefficients.  
The model in table [\ref{fnyvariant}] is a variant
of the model of ref. \cite{fny}. This model utilizes asymmetric
boundary conditions. There are three complexified internal fermions, which 
are the same as those of
the model of table [\ref{fnymodel}], and the pairing of the left--moving
world--sheet fermions is identical. In terms of its phenomenological 
characteristics the model is a three generation model, each arising
from the sectors $b_1$, $b_2$ and $b_3$. The four dimensional gauge
group is the same as that of ref. \cite{fny}. The model utilizes
the doublet--triplet splitting mechanism of ref. \cite{ps}, which
arises from the asymmetric boundary condition assignments in the basis
vectors that break the observable $SO(10)\rightarrow SO(6)\times SO(4)$.
Similarly, the model yields tri--level Yukawa couplings to the three
$+2/3$ charged quarks, but not to the  $-1/3$ charged quarks, which
is a result of the asymmetric boundary condition assignment in the basis
vector that breaks $SO(10)\rightarrow SU(5)\times U(1)$.
In the model of table 
[\ref{fnyvariant}] all the untwisted moduli are left
in the massless spectrum. It is instructive to rewrite the boundary
conditions of the internal fermions in this model in the notation of 
table [\ref{symmetricthirring}]
\begin{eqnarray}
     &\begin{tabular}{c|cccc|cccc|cccc}
 ~&   $y^1{\omega}^1$
  &   ${\bar y}^1{\bar\omega}^1$
  &   $y^2{\omega}^2$
  &   ${\bar y}^2{\bar\omega}^2$
  &   $y^3{\omega}^3$
  &   ${\bar y}^3{\bar\omega}^3$
  &   $y^4 {\omega}^4$
  &   ${\bar y}^4{\bar\omega}^4$
  &   $y^5{\omega}^5$
  &   ${\bar y}^5{\bar\omega}^5$
  &   $y^6{\omega}^6$
  &   ${\bar y}^6{\bar\omega}^6$ \\
\hline
\hline
$\alpha$ & 11 & 00 & 00 & 00  & 11 & 00 & 00 & 00  & 00 & 00 & 11 & 00
\\
$\beta$  & 00 & 11 & 00 & 00  & 00 & 11 & 11 & 11  & 00 & 00 & 00 & 11
\\
$\gamma$ & 00 & 11 & 00 & 00  & 00 & 11 & 00 & 00  & 11 & 11 & 00 & 11
\\\nonumber
\end{tabular}\\
\label{internalfnyvariant}
\end{eqnarray}
In this notation it is apparent that despite the utilization of asymmetric
boundary conditions, the specific pairing of world--sheet fermions
allows the retention of all the untwisted moduli. 
In terms of the bosonized
variables the boundary conditions translate to
\beq
     \begin{tabular}{c|c|c|c}
      &   $Z_1$     &      $Z_2$   &      $Z_3$  \\
\hline
\hline
$\alpha$ &
                $Z_1+(1+i2)\pi$ &
                $Z_2+(1+i2)\pi$ & $Z_3+(2+i)\pi$ \\
$\beta$  &
                $Z_1+(1+i2)\pi$ &
                $Z_2+2\pi$      & $Z_3+(2+i)\pi$\\
$\gamma$ &
                $Z_1+(1+i2)\pi$ &
                $Z_2+(2+i2)\pi$ & $Z_3+i\pi$ 
\end{tabular}
\label{fnyvcomfercombc123}
\eeq
{}From eq. (\ref{fnyvcomfercombc123}) it is evident
that all
three complex planes retain the complex geometry, and hence
the three K\"ahler and three complex moduli remain in the spectrum.
In this model the reduction to three generations is attained
solely by the shift identifications in the real and complex 
directions, which are asymmetric in terms of the fermionic
boundary conditions, but symmetric in terms of the bosonic
variables. 

The investigation above of the moduli in the three generation
free fermionic models is, of course, not exhaustive, but rather
illustrative.
The moduli of other models in this class may be similarly investigated.
I give here a cursory view of several additional models. 
As illustrated above the determinantal factor in regard to the
untwisted moduli is the pairing of the left-- and right--moving
fermions from the set $\{y,\omega|{\bar y},{\bar\omega}\}^{1,\cdots,6}$.
In the case of symmetric boundary conditions, with 12 left--moving real 
fermions combined with 12 real right--moving fermions to form 12
Ising model operators, the moduli fields always remain in the spectrum
and the six compactified coordinates maintain their geometrical
character. When the real fermions are combined to form
complex fermions the situation is more varied, as illustrated above.

The determination of the moduli, however, does not depend on the choice
of the observable four dimensional gauge group. To exemplify that I consider
the model of ref. \cite{alr}, which utilizes the same complexification
of the real fermions as that of the model of table
[\ref{model2}]. It is then found
that the retained and projected moduli in this model are those in eqs.
(\ref{retmodmod2}) and (\ref{promodmod2}), respectively. Of course,
there may be a correlation between the assignment of boundary
conditions to the real fermions $\{y,\omega\}_{L,R}$ and those
that determine the four dimensional gauge group. Such dependence
may arise because of the modular invariance constraints \cite{fff}. 
But typically we
may relegate this correlation to the hidden sector, and therefore
the observable sector is not affected. 

In the case of three complex right--moving fermions, and their
corresponding left--moving complexified fermions, each pair being
associated with a distinct complex planes, we noted in eqs.
(\ref{m278pairing}) and (\ref{fnypairing}),
two cases of pairings, with differing consequences for
the untwisted moduli fields. The case of (\ref{fnypairing})
was further investigated in the model of table
[\ref{fnyvariant}]. The moduli with the pairing of eq.
(\ref{m278pairing}) may be similarly investigated
in models that utilize this pairing. An example of 
such a model is the model in table [\ref{m274}], 
\begin{eqnarray}
 &\begin{tabular}{c|c|ccc|c|ccc|c}
 ~ & $\psi^\mu$ & $\chi^{12}$ & $\chi^{34}$ & $\chi^{56}$ &
        $\bar{\psi}^{1,...,5} $ &
        $\bar{\eta}^1 $&
        $\bar{\eta}^2 $&
        $\bar{\eta}^3 $&
        $\bar{\phi}^{1,...,8} $\\
\hline
\hline
  ${\alpha}$  &  0 & 0&0&0 & 1~1~1~0~0 & 0 & 0 & 0 & 1~1~1~1~0~0~0~0 \\
  ${\beta}$   &  0 & 0&0&0 & 1~1~1~0~0 & 0 & 0 & 0 & 1~1~1~1~0~0~0~0 \\
  ${\gamma}$  &  0 & 0&0&0 &
		${1\over2}$~${1\over2}$~${1\over2}$~${1\over2}$~${1\over2}$
	      & ${1\over2}$ & ${1\over2}$ & ${1\over2}$ &
                ${1\over2}$~0~1~1~${1\over2}$~${1\over2}$~${1\over2}$~0 \\
\end{tabular}
   \nonumber\\
   ~  &  ~ \nonumber\\
   ~  &  ~ \nonumber\\
     &\begin{tabular}{c|c|c|c}
 ~&   $y^3{y}^6$
      $y^4{\bar y}^4$
      $y^5{\bar y}^5$
      ${\bar y}^3{\bar y}^6$
  &   $y^1{\omega}^5$
      $y^2{\bar y}^2$
      $\omega^6{\bar\omega}^6$
      ${\bar y}^1{\bar\omega}^5$
  &   $\omega^2{\omega}^4$
      $\omega^1{\bar\omega}^1$
      $\omega^3{\bar\omega}^3$
      ${\bar\omega}^2{\bar\omega}^4$ \\
\hline
\hline
$\alpha$ & 1 ~~~ 1 ~~~ 1 ~~~ 0  & 1 ~~~ 1 ~~~ 1 ~~~ 0  & 1 ~~~ 1 ~~~ 1 ~~~ 0
\\
$\beta$  & 0 ~~~ 1 ~~~ 0 ~~~ 1  & 0 ~~~ 1 ~~~ 0 ~~~ 1  & 1 ~~~ 0 ~~~ 0 ~~~ 0
\\
$\gamma$ & 0 ~~~ 0 ~~~ 1 ~~~ 1  & 1 ~~~ 0 ~~~ 0 ~~~ 0  & 0 ~~~ 1 ~~~ 0 ~~~ 1
\\
\end{tabular}
\label{m274}
\end{eqnarray}
with a suitable choice of GSO projection coefficients.
This model is published in \cite{top}, and has similar
features to the model of table [\ref{m278}], with the difference being that
[\ref{m274}] yields bottom--quark and tau--lepton
Yukawa couplings at the quartic level of the superpotential \cite{top},
whereas [\ref{m278}] yields such couplings only at the quintic
order \cite{eu}. In the model of table. [\ref{m274}] one finds again that all
the untwisted moduli fields are projected out. Another example 
of a model in this class is the model on page 14 of \cite{pairings},
which utilizes the pairings of eq. (\ref{m278pairing}).
This model is a variation of the model of table [\ref{m274}], with
the boundary conditions in the vector $\gamma$ chosen
such that the sector $b_2$ produces trilevel bottom--type
Yukawa coupling rather than top--type. Again in this model
all the untwisted moduli are projected out. In fact, we
may conjecture that in general in free fermionic models
that utilize the NAHE--set of boundary condition basis vectors,
and the pairing of eq. (\ref{m278pairing}), the moduli are always
projected out. The reason is that in order to reduce the
number of generations to three, one from each of the twisted
sectors, $b_1$, $b_2$ and $b_3$, we have to break the degeneracy
between the complexified left--moving and right--moving fermions
in each sector. This entails the assignment of asymmetric boundary
conditions with respect to these complexified fermions, in at least
one of the basis vectors, $\alpha$, $\beta$ or $\gamma$. In
the case of the pairing of eq. (\ref{m278pairing}) this leads
to the projection of all the untwisted moduli, as noted above,
because of the fact that this pairing mixes all the six
left--moving triplets of the $SU(2)^6$ automorphism algebra. 

\section{Twisted moduli}

I now turn to show that is the class of three generation
string models under consideration here
moduli which arise from the twisted sectors are also
projected from the massless spectrum. 
To see how this comes about we start with the set of basis vectors 
$\{1,S,\xi_1,\xi_2,b_1,b_2\}$ \cite{foc}. This set of basis 
vectors generates a model with $E_6\times U(1)^2\times E_8\times SO(4)^3$,
with 24 matter states in the 27 representation of the observable $E_6$ 
gauge group, arising from the twisted sectors. These states are decomposed 
in the following way under $E_6\rightarrow SO(10)$. The spinorial 16
representations
of $SO(10)$ arise from the sectors $b_1$, $b_2$ and $b_3$,
whereas the vectorial
10 representations arise from the sectors $b_j+\xi_1$. Here,
the basis vector $\xi_1$ 
produces the space-time vector gauge bosons that enhance
the $N=4$ observable $SO(16)$
gauge group to $E_8$. In addition to the vectorial 10
representation the sectors $b_j+\xi_1$,
also produce a pair of $SO(10)$ singlets, one of which
is embedded in the $27$ of 
$E_6$, whereas the second is indentified with a twisted moduli.
Therefore, the models 
with an $E_6$ observable gauge group contain additional 24
twisted moduli, which matches
the number of chiral matter states in the model, as it should. 

In the realistic free fermionic models the observable gauge
group is broken from $E_6$ 
to $SO(10)\times U(1)$. This can be achieved in two
equivalent ways. One possibility is
to replace the vector $\xi_1$ with the vector $2\gamma$.
In this case
the gauge group of the $N=4$ vacuum, generated by the subset of
basis vectors $\{1,S,2\gamma,\xi_2\}$,
is $SO(12)\times SO(16)\times SO(16)$. The space-time vector
bosons of the four dimensional 
gauge group in this case are obtained from the NS sector,
the sector $\xi_2$, and the sector 
$\xi_2+2\gamma$. An equivalent way to produce the same $N=4$
string vacuum is to break the $E_8\times E_8$
gauge group by the choice of GSO phase eq. (\ref{gsophasexi1xi2})
in the one--loop string partition function. 
One choice of this phase preserves the states from the
sectors $\xi_1$ and $\xi_2$, and therefore  enhances
the NS $SO(16)\times SO(16)$ gauge symmetry to $E_8\times E_8$.
The second choice projects out those states. 
In the second case the sectors $b_j+2\gamma$, or
$b_j+\xi_1$ produce states in the 16 vectorial
representation of the hidden $SO(16)$ gauge group, whereas
the vectorial states in the $(5+{\bar 5})$
and $(1+{\bar 1})$, of the observable $SO(10)$ group,
generated by the world--sheet fermions
${\bar\psi}^{1,\cdots,5}$ are projected out from the spectrum.
Hence, in this case all
the moduli that arise from twisted sectors are projected out. 
The only states that arise from the twisted sectors in three
generation models are observable or hidden matter states. 

\section{Discussion and conclusions}

In this paper I investigated the untwisted moduli
fields in the three generation free fermionic heterotic--string models.
This class of string vacua produced some of the most
realistic string models constructed to date. Not only does
it produce three generations of chiral fermions under the 
Standard--Model gauge group with potentially phenomenologically
viable couplings to the electroweak Higgs doublet fields,
but it also affords the attractive embedding of the Standard
Model spectrum in $SO(10)$ representations. This property
ensures the canonical GUT normalization of the weak hypercharge,
which in turn facilitates the agreement of the heterotic--string
coupling unification with the experimental data. Thus, these models
produce a gross structure, which is compelling from a phenomenological
point of view. It ought to be emphasized that one should not regard
any of the existing models as providing a completely realistic
phenomenology, but merely as providing a probe into what may be
some of the ingredients of the eventually true vacuum. From this
perspective, a vital property of the free fermionic models is their
connection to $Z_2\times Z_2$ orbifold compactifications. 

The free fermionic formalism facilitated the construction of the
three generation models and the analysis of some of their phenomenological
characteristics. However, this method is formulated, a priori, at a fixed
point in the moduli space and the immediate notion of the underlying
geometry of the six dimensional compactified manifold is lost.
In particular, the identification of the untwisted moduli fields, and
their role in the low energy effective theory, is encumbered.
These are reincorporated into the models by identifying the moduli
fields as the coefficients of the Abelian Thirring interactions
\cite{lny}. The moduli fields in the free fermionic models are 
therefore in correspondence with the Abelian Thirring terms that are
invariant under the GSO projections, induced by the basis vectors
that define the models. 

In this paper the issue of untwisted moduli in the realistic three generation
free fermionic models was investigated. The existence of models in which
all the untwisted K\"ahler and complex structure moduli are projected out by
the generalized GSO projections was demonstrated. The conditions
for the projection of all the moduli were identified, and compared
to other similar models in which the untwisted moduli are retained. 
The basic condition that enable the projection of all the untwisted 
moduli is when the fermionic boundary conditions are such that they
correspond to left--right asymmetric boundary conditions with respect
to all the six real coordinates of the six dimensional internal manifold.
Additionally, it was shown that in this class of three generation models
the $E_6$ symmetry is broken to $SO(10)\times U(1)$ by a GSO projection.
As a consequence all the moduli that arise from the twisted sectors
are also projected out in these models. 

The existence of quasi--realistic models in which all the
untwisted K\"ahler and complex structure moduli are projected out is
fascinating. It offers a novel perspective on the existence of extra
dimensions in string theory, and on the problem of moduli stabilization.
The untwisted moduli are those that govern
the underlying geometry, and hence the physics of the extra dimensions.
Reparameterization invariance in string theory introduces the need
for additional degrees of freedom, beyond the four space--time dimensions,
to maintain the classical symmetry in the quantized theory. These additional
degrees of freedom may be interpreted as extra dimensions, which are 
compactified and hidden from contemporary experimental observations.
Thus, the consistency of string theory gives rise to the notion of
extra dimensions, which in every other respect is problematic.
In particular, it raises the issue of what is the mechanism that
selects and fixes the parameters of the compactified space.
However, if there exist string models in which all the untwisted
moduli are projected out by the GSO projections, it means that in these
vacua the parameters of the extra dimensions are frozen. In fact, in these
string vacua the extra degrees of freedom needed for consistency
cannot be interpreted as extra dimensions, as it is not possible 
to deform from their fixed values, and there is no notion of a continuous
classical geometry.

The problem of moduli stabilization in string theory attracted considerable
attention in the literature \cite{recentpaper}. Most of the studies have
been directed toward stabilization of the extra dimensions in the effective
low energy field theory that emerges from the underlying string theory. 
The primary obstacle to this is the fact that there are no potential
terms for the moduli fields to all orders in perturbation theory.
One then, in general, has to rely on the appearance of nonperturbative
potential terms, or the utilization of internal fluxes \cite{recentpaper}.
However, the mechanism that fixes the moduli in the free fermionic models
is an intrinsically string theoretic mechanism. The reason is that this
mechanism utilizes asymmetric boundary conditions. The possibility to
separate the internal dimensions into left-- and right--movers, 
and to assign different transformation properties to them,
is intrinsically string theoretic and is nonsensical in 
the effective low energy point quantum field theories,
considered to date. 

String theory provides a consistent framework for perturbative
quantum gravity. In this context it provides the quantum, albeit
perturbative, probe of the underlying geometry. The effective
low energy point quantum field theory, on the other hand,
treats the underlying geometry as a classical geometry. 
The existence of quasi--realistic string vacua in which all the
untwisted  moduli are fixed at the string level may tell us
that, although string theory requires the additional degrees of freedom 
for its consistency, these degrees of freedom are not necessarily
realized as extra continuous classical dimensions, in the
phenomenologically viable cases. These vacua live
intrinsically in four space--time dimensions, and there is no notion
of extra geometrical dimensions in the low energy effective point quantum
field theory. Thus, while the geometrical notion of the extra degrees
of freedom provides a useful mean to classify the string vacua,
these do not have a physical realization.
This situation is similar to the way in which the GUT symmetries
are broken in the quasi--realistic free fermionic heterotic--string models
by the generalized GSO projections.
Namely, also in this case, while the Standard Model states fall into
representations of the underlying $SO(10)$ symmetry, the GUT group is
broken directly at the string level, and is 
not realized as a gauge symmetry in the effective low energy point
quantum field theory. 

The existence of quasi--realistic string vacua in which all the untwisted
moduli are projected out by the GSO projections is fascinating and intriguing.
As discussed in this paper, such string vacua exist among the so--called
realistic free fermionic models. We may then ask what are the properties
of these models that enabled this outcome, and whether it is unique
to this class of models. It should be emphasized that although
the free fermionic formalism is formulated at a fixed point in the
moduli space, this does not yet entail the absence of an underlying
geometry, and the geometrical degrees of freedom are reincorporated 
in the form of the Abelian world--sheet Thirring interactions.
The string vacua in which all the untwisted moduli are fixed
represent a special subclass of the three generation models,
and the projection of the moduli is highly correlated with the
reduction of the number of generations. That is, it is only in the
special case of the three generation models that one may expect to find
vacua in which all the untwisted moduli are projected out.

The defining property of the three generation free fermionic
heterotic string models is their relation to $Z_2\times Z_2$
orbifold compactifications \cite{foc}. We may therefore ask whether the
$Z_2$ projections, and the $Z_2\times Z_2$ orbifold possess
some special properties that enables the projection of all the
untwisted geometrical moduli, and distinguishes it from other
compactifications. Naturally, an answer to this question requires
further investigation. However, we may note that the special
property of the $Z_2\times Z_2$ orbifold is that it may act on
the internal dimensions as real, rather than complex, dimensions.
As discussed in section \ref{min3gm} it is this property that enables
the projection of all the untwisted K\"ahler and complex structure moduli
in the free fermionic models that utilize the pairings of eq.
(\ref{m278pairing}). Whether or not similar results may be
obtained in other classes of string compactifications
is an interesting question that requires further research.

It should be emphasized that the results of this paper
do not imply a complete solution to the moduli problem
in string theory. A moduli field, in general, is a field that
does not have a potential to all orders in perturbation theory,
and therefore its vacuum expectation value is unconstrained,
and it does not get a mass term. The string models contain other sources
of moduli fields, aside from the perturbative untwisted geometrical
moduli, and the twisted sectors moduli.
These include: the dilaton field;
and the possibility of charged moduli. Furthermore, 
the class of vacua under consideration here are supersymmetric
and there may exist moduli associated with the supersymmetric
flat directions. Moreover, in string theory, each continuous coupling
naively implies the existence of a moduli field, and therefore
prior to fixing all the couplings in a given model, it seems
premature to claim that the entire issue of moduli has been addressed.
The untwisted moduli fields are, however, those that parameterize
the shape and size of the underlying six dimensional compactified space.

In regard to the dilaton field, the models investigated here
are perturbative heterotic strings. Following the string duality
advances, we now know that the heterotic string is a perturbative
limit of the more fundamental quantum M--theory. At present
we have no knowledge about this quantum theory, aside from the
existence of its effective limits. The heterotic limit is an
expansion in vanishing string coupling, and therefore one would
not expect to be able to fix the dilaton VEV in this limit 
(although one may entertain some nonperturbative possibilities
\cite{dilaton}). 
In this paper it was found that fixation of
the other untwisted moduli is achieved by an intrinsic string mechanism,
and hence at the perturbative quantum gravity level.
However, at present the quantum
M--theory is not available. The lesson from the current paper is
that the quantum M--theory may allow further possibilities to the 
problem of dilaton stabilization, which are not readily gleaned in the
effective low energy point quantum field theory description.

Additional flat direction moduli
and charged moduli may, in general, exist in the
string models. Their existence, or absence,
in the string model is more model dependent and requires
a model by model analysis. However, the structure of the
three generation free fermionic models suggests that flat,
or charged, moduli are not interchanged with untwisted moduli,
and hence an underlying continuous geometrical manifold is
not restored.

Finally, it should be emphasized that whether or not extra dimensions
play a physical role in nature would, of course, require further 
study and investigation. From the discussion in sections \ref{rffm} and 
\ref{moduliinffm} it is found that
the untwisted sector of the NAHE--based free fermionic models produces
three complex and three K\"ahler structure moduli.
This outcome remains valid in any free fermionic model which is
left--right symmetric. As discussed in section \ref{rffm}
there do exist three generation free fermionic models that 
are left--right symmetric. Therefore in these models the
entire set of moduli fields exist in the effective low energy
field theory and the moduli remain unfixed. 
However, the important point is that the free fermionic models
also allow for boundary conditions which are left--right asymmetric.
Naturally, the space of models is vast and we can construct models
in which asymmetric boundary conditions are assigned
on one, two or three of the complex planes. The remarkable fact
is the existence of three generation models in which asymmetric boundary
conditions are assigned to all three complex planes.
In this set of models the entire set of untwisted moduli are projected
out by the GSO projections and the untwisted moduli are fixed.
The results of this paper indicate
the existence of quasi--realistic string vacua in which the extra
dimensions do not possess a classical physical realization.
On the other hand there are also three generation models in 
which the moduli are retained and the geometrical description is
maintained, and models in which some of the extra dimensions are
frozen whereas others are undetermined.
Which, if any, of these possibilities is relevant to nature,
would, of course, remain, for the time being, a hotly contested issue.
Moreover, the models may of course still contain additional moduli and
the role of these moduli requires more detailed investigation.
We should also remember that we are discussing
here perturbative heterotic string models. In this limit the
dilaton remains unfixed, and it may be that nonperturbative effects
may give rise to additional moduli. 
Nevertheless, it is extremely intriguing that in the class of three
generation free fermionic models, that are constructed in the 
vicinity of the self--dual point under the T--duality transformations,
one finds that the models possess the intrinsic mechanism to fix 
all the untwisted geometrical moduli. One would anticipate that the
self--dual point, being the symmetry point under the T--duality
transformations, plays a vital role in the vacuum selection \cite{selfdual}.
The three generation free fermionic models then highlight the class of
$Z_2\times Z_2$ orbifolds as the naturally relevant one.
The availability, in this class, to act asymmetrically on all
the six real compactified dimensions, then affords the possibility
of fixing all the untwisted
K\"ahler and complex structure moduli in these models.

\bigskip
\section{Acknowledgments}

This work is supported in part by PPARC. I would like to thank the
anonymous referee for valuable comments. 


\medskip

\bibliographystyle{unsrt}

\end{document}